\documentclass[%
jcp,%
 amsmath,amssymb,
 reprint,%
onecolumn,
]{revtex4-1}

\usepackage{graphicx}
\usepackage{dcolumn}
\usepackage{bm}

\begin{document}

\preprint{AIP/123-QED}

\title{Structural and electrostatic properties between pH-responsive polyelectrolyte brushes studied by augmented strong stretching theory}

\author{Jun-Sik Sin}

 \email{js.sin@ryongnamsan.edu.kp}
  \affiliation{Natural Science Center, Kim Il Sung University, Taesong District, Pyongyang, Democratic People's Republic of Korea}


\begin{abstract}
\large
In this paper, we study electrostatic and structural properties between pH-responsive polyelectrolyte brushes by using a strong stretching theory accounting for excluded volume interactions, the density of polyelectrolyte chargeable sites and the Born energy difference between the inside and outside of the brush layer.

In a free energy framework, we obtain self-consistent field equations to determine electrostatic properties between two pH-responsive polyelectrolyte brushes. We elucidate that in the region between two pH-responsive polyelectrolyte brushes, electrostatic potential at the centerline and osmotic pressure increase not only with excluded volume interaction, but also with density of chargeable sites on a polyelectrolyte molecule.

 Importantly, we clarify that when two pH-responsive polyelectrolyte brushes approach each other, the brush thickness becomes short and that a large excluded volume interaction and a large density of chargeable sites yield the enhanced contract of polyelectrolyte brushes. In addition, we also demonstrate how the influence of such quantities as pH, the number of Kuhn monomers, the density of charged sites, the lateral separation between adjacent polyelectrolyte brushes, Kuhn length on the electrostatic and structural properties between the two polyelectrolyte brushes is affected by the exclusion volume interaction. 

Finally, we investigate the influence of Born energy difference on the thickness of polyelectrolyte brushes and the osmotic pressure between two pH-responsive polyelectrolyte brushes.

\end{abstract}
\maketitle

\large
\section{\label{sec:level1}Introduction}
Grafting polyelectrolyte brushes onto solid-liquid interfaces is a promising strategy to fabricate   functionalized nanochannels and nanoparticles because it is capable to regulate many physicochemical properties of interfaces formed between hard surfaces and ionic solutions. \cite%
{ali_jacs_2008,vilozny_ns_2013,sadeghi_aca_2020,ali_jacs_2009,sadeghi_pccp_2021,kreer_sm_2016,nap_jcp_2018,marins_acs_2018,kegler_prl_2007,sachar_matter_2020,zhao_lang_2018}

Ionization and charging of strong polyelectrolytes depends on only the ionic strength of the electrolyte solution, whereas weak (or annealed) polyelectrolytes involve pH-dependent ionization and charging and hence are widely employed to receive a variation in pH or ionic concentration. 

For weak polyelectrolyte brushes, a change in solution pH or ionic strength causes noticeable changes in its conformation or its brush thickness which is the reason why weak polyelectrolyte brushes not only is called pH-responsive brushes but also are widely employed for industrial applications such as biosensing\cite%
{ali_jacs_2008}, current rectifiers \cite%
{vilozny_ns_2013, sadeghi_aca_2020, ali_jacs_2009} and power generation in nanochannels \cite%
{sadeghi_pccp_2021}, polymer-brush lubrication \cite%
{kreer_sm_2016}, nanoparticles for biomedical applications \cite%
{nap_jcp_2018,marins_acs_2018, kegler_prl_2007}, nanofiltration \cite%
{sachar_matter_2020} and  many more \cite%
{zhao_lang_2018, zhulina_jcp_2002}

Some studies  to describe pH-responsive PE brushes\cite%
{pincus_macromolecules_1991, ross_macromolecules_1992, borisov_jp_1991, joanny_macromolecules_1993, borisov_macromolecules_1994, zhulina_macromolecules_1995, zhulina_sm_2012, zhulina_macromolecules_1994a} aimed to obtain scaling laws by considering the balance between the different energies (elastic, electrostatic, and excluded volume) to yield the brush thickness as functions of quantities such as the grafting density and charge density of the brushes, number of monomers, and the concentration of the added salt. For establishing a more complete picture of pH-responsive polyelectrolyte brushes, the Poisson-Boltzmann equation \cite%
{das_jpcb_2015,das_rsc_2015,das_jpcb_2016,das_jpcb_2017,das_sm_2018,das_jpcb_2018,biesheuvel_macromolecules_2008} was widely studied for the electrostatics of the induced electric double layer (EDL)by using the monomer interactions such as the Alexander-de-Gennes model and the parabolic model. Moreover, a more innovative approach for pH-responsive polyelectrolyte brushes is the strong stretching theory \cite%
{zhulina_jcp_1997,zhulina_macromolecules_1995a,zhulina_jcp_2017b,zhulina_macromolecules_2000,zhulina_lang_2011} presented to describe the polyelectrolyte brushes. 

The self-consistent strong stretching theory and the Poisson-Boltzmann equation have also been employed to study the configuration (or brush thickness) of two interacting pH-responsive polyelectrolyte brushes and proposed that when approaching two interacting polyelectrolyte brushes, the thickness of polyelectrolyte brush can be reduced.

Unfortunately, the papers \cite%
{zhulina_jcp_1997, zhulina_macromolecules_1995a, zhulina_jcp_2017b, zhulina_macromolecules_2000, zhulina_lang_2011} discard the excluded volume interaction between monomer units in a polyelectrolyte chain and hence considers the polyelectrolyte brushes being in a $\theta$ -solvent. 
In many realistic experiments \cite%
{kumar_jps_2016, wu_macromolecules_2007, ito_jacs_1997}, physicochemical and structural properties of pH-responsive PE brushes were performed by using good solvents and, accordingly,  the theory of \cite%
{zhulina_jcp_1997} dealing with $\theta$ -solvent doesn't apply to such cases. On the other hand, since the theory does not account for the density of chargeable sites of the polyelectrolyte brushes, the theory is unable to study the variation of polyelectrolyte brush thickness with the density of chargeable sites of the polyelectrolyte brush, as reported in \cite%
{zhulina_macromolecules_2014, kumar_macromolecules_2005, naji_epje_2003}. 

In order to account for such effects, the authors of \cite%
{das_sm_2019a} proposed a self-consistent field approach to describe the behavior of the pH-responsive brush by considering (a) the excluded volume interactions between the polyelectrolyte segments and (b) a more expanded form of the mass action law valid for $\gamma a^3$. The study unravels an increase of the brush thickness due to the excluded volume interaction which is originated from polyelectrolyte inter-segmental repulsion, as well as a variation in the brush thickness for $\gamma a^3$, attributed to counterion-induced osmotic swelling of the brushes. Furthermore, with the help of the augmented strong stretching theory, the authors provided the fascinating theoretical analysis \cite%
{das_sm_2019b, das_ep_2020, das_pf_2020, das_pre_2020, das_jfm_2021} of liquid transport driven by an axial temperature gradient, concentration gradient, electric potential gradient in polyelectrolyte-brush-grafted nanofluidic channel.
However, to the best of our knowledge, there is no researcher which discusses how the excluded volume interaction and the expanded form of the mass action law affect the electrostatic interaction and structure between pH-responsive polyelectrolyte brushes.

To unravel the issue, we generalize the augmented strong stretching theory for a pH-responsive polyelectrolyte brush presented in \cite%
{das_sm_2019a} to the case of two interacting pH-responsive polyelectrolyte brushes and investigate the influence of the excluded volume interaction and the expanded form of the mass action law on interaction and structural properties between two pH-responsive brushes in connection with other quantities.

Here, it is worthwhile to provide a brief overview of the modeling studies of electrostatic interactions between two polymer-grafted-surfaces.  In fact, although many works \cite%
{duval_lang_2011, duval_lang_2015, ohshima_lang_2006, Das_csb_2018,  dahnert_jcis_1994, ohshima_jcis_2002, richmond_jcsf_1974, richmond_jcsf_1975, nelson_ijtb_1975, micklavic_jpc_1994, mickavic_jcis_1999, ohshima_cps_2014, das_rsc_2015b, duval_pccp_2011, sin_csa_2021} have dealt with the electrostatic interaction between the polyelectrolyte grafted particles in electrolyte solutions, most of them are based on the assumption that lengths of chains in a polyelectrolyte layer are equal and constant and are independent of external stimuli such as temperature, pH and salt concentration in electrolyte solutions. 
Consequently, such an oversimplification allows the theories not to predict the influence of a polyelectrolyte chain on properties of polyelectrolyte grafted particles. Moreover, recent works \cite%
{sin_csa_2021, horno_jcis_2003, sadeghi_eleccom_2017, gopmandal_pre_2018, sadeghi_csb_2018, gopmandal_pre_2020, gopmandal_ep_2021, poddar_sm_2016, pandey_csa_2021} anticipate that physicochemical properties of polyelectrolyte grafted nanochannels or nanoparticles is significantly affected by the Born energy difference between the inside and outside of polyelectrolyte layer.

Thus we aim at investigating the electrostatic and structural properties between two interacting pH-responsive polyelectrolyte brushes. 
The rest of the paper is organized as follows: We will first consider thermodynamics of the system consisting of two interacting planar polyelectrolyte brushes in an electrolyte solution and then derive the electrostatic potential distribution and structural properties of pH-responsive polyelectrolyte brushes tethered on neutral surfaces.  Finally, the influence of various quantities on brush thickness, electrostatic potential and osmotic pressure are demonstrated in detail. 

\section{Theory}

We consider two parallel pH-responsive polyelectrolyte brushes separated by a distance $2h \left(-h < x < h \right)$. Theoretical description of the brushes is based on the free energy formalism using the augmented strong stretching theory accounting for a more generic mass action law and the effects of the polyelectrolyte excluded volume interactions. 

The present theory will present the equilibrium configuration of the polyelectrolyte brushes and the equilibrium electrostatics of the brush-induced electric double layer by minimizing the total free energy of a given brush molecule that consists of the elastic, excluded volume, electrostatic, and ionization energies of the brush and the electrostatic energy of the brush-induced electric double layer. 
\begin{figure}
\begin{center}
\includegraphics[width=0.8\textwidth]{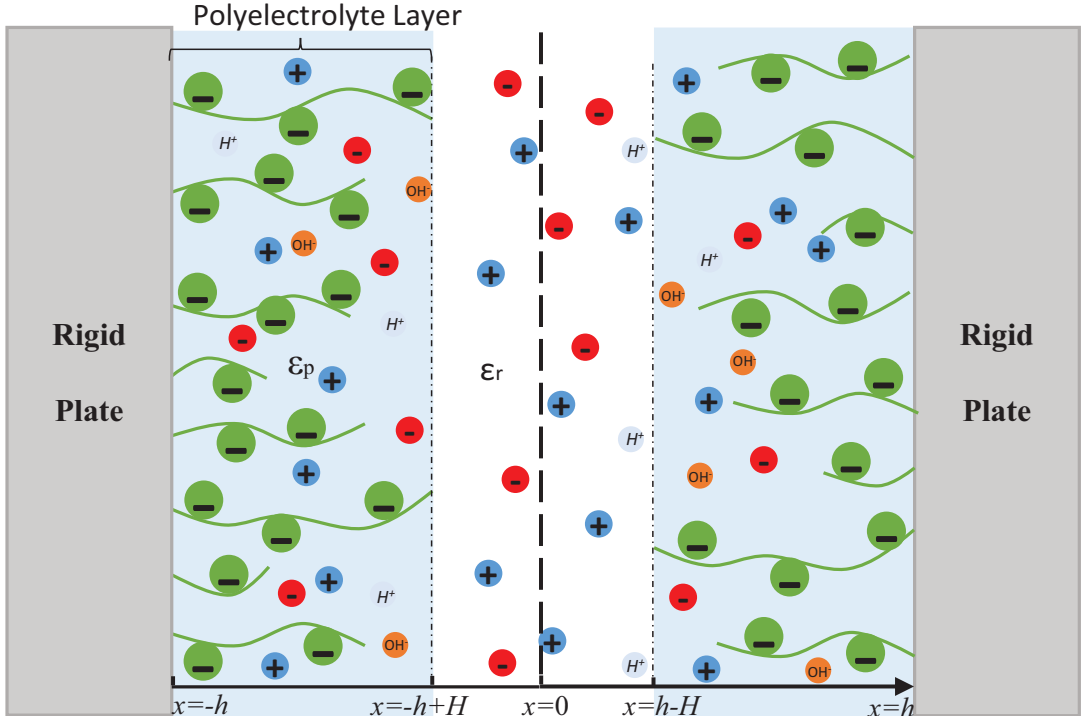}
\caption{(Color online) Schematic of two pH-responsive polyelectrolyte brushes in an electrolyte solution. Here the polyelectrolyte ions are represented in green, electrolytic anions and cations are represented in red and blue, respectively. Hydrogen ions and hydroxyl ions are represented in cyan and orange.
}
\label{fig:1}
\end{center}
\end{figure}

The total energy functional $F$ of the PE brush system consists of the elastic $F_{els}$, excluded volume $F_{EV}$, electrostatic $F_{elec}$, and ionization $F_{ion}$ free energies of a polyelectrolyte brush molecule, the electrostatic energy of the EDL ($F_{EDL}$) induced by the brush and the free energy due to Born energy difference($F_{Born}$) . 
\begin{eqnarray}
F = F_{els}  + F_{EV}  + F_{elec}  + F_{ion}  + F_{EDL}+ F_{Born}
\label{eq:1}.
\end{eqnarray}
Due to axial symmetry of the present system, we shall only consider the left half of the region between two interacting pH-responsive brushes in our subsequent calculation. 
In the same way as in \cite%
{zhulina_jcp_1997, das_sm_2019a}, we can express
\begin{eqnarray}
F_{els} = \frac{3k_B T}{{2pa^2 }}\int_{ - h}^{ - h + H} {g\left( {x'} \right)dx'} \int_{ - h}^{x'} {E\left( {x,x'} \right)dx} 
\label{eq:2},
\end{eqnarray}
\begin{eqnarray}
F_{EV} = \frac{k_B T}{{\sigma a^3 }}\int_{ - h}^{ - h + H} {f_{EV} \left[ {\phi \left( x \right)} \right]dx} 
\label{eq:3}.
\end{eqnarray}
where $k_B, T, p, a, H, \sigma$ and $ l_0$ denote the Boltzmann constant, the temperature, chain rigidity, Kuhn length, brush thickness, polyelectrolyte brush grafting density, and the lateral separation between the adjacent polyelectrolyte brushes, respectively. Here $\sigma \propto \frac{1}{l_0^2}$.
$\phi \left( x \right)$  and $f_{EV} \left[ {\phi \left( x \right)} \right]\left( { \approx v\phi ^2  + \omega \phi ^3 } \right)$
  represent the dimensionless monomer distribution profile of a polyelectrolyte chain and the non-dimensionalized per unit volume free energy associated with the excluded volume interactions, respectively.  $E\left( {x,x'} \right) = \frac{{dx}}{{dn}}$ expresses chain stretching (for a chain whose end is at $x'$) at a location $x$ and $g\left(x'\right)$(the normalized chain end distribution function) is expressed as
\begin{eqnarray}
\int_{ - h}^{ - h + H} {g\left( {x'} \right)dx'}  = 1
\label{eq:4}.
\end{eqnarray}
The Born energy difference of the $i$-{\it{th}} ionic species , $\Delta W_i$, is expressed as follows
\begin{equation}
\Delta W_i =\frac{{e}^2}{8\pi\varepsilon_0 r_i}\left(\frac{1}{\varepsilon_p}-\frac{1}{\varepsilon_r}\right)=\frac{{e}^2}{8\pi\varepsilon_0 \varepsilon_r r_i}\left(\frac{1}{\zeta}-1\right)
\label{eq:51}
\end{equation}
where $r_i\left(i = 1, 2\right)$ is the radius of ion having the elementary charge $e$. $\varepsilon_0$  is the absolute permittivity of vaccum,  $\varepsilon_p$  and $\varepsilon_r$  are the relative permittivities of the inside and outside of polyelectrolyte brush layer, respectively. $\zeta=\frac{\varepsilon_p}{\varepsilon_r}$. For simplicity, we assume that anions and cations have an equal value of hydrated radius, $r$. Eq. (\ref{eq:51}) imply that the energy difference decreases with the permittivity of polyelectrolyte layer. If the polyelectrolyte brush layer contains less water, the difference in permittivity between inside and outside of polyelectrolyte brush layer is enhanced and finally the Born energy difference will increase.
\begin{equation}
F_{Born}  = \int\limits_{ - h}^{ - h + H} {\Delta W\left( {n_ +   + n_ -   + n_{H^ +  }  + n_{OH^ -  } } \right)dx} 
\label{eq:A5}
\end{equation}
Next, following  \cite%
{das_sm_2019a}, $F_{elec}  + F_{EDL}$  can be expressed as 
\begin{eqnarray}
\begin{array}{l}
 F_{elec}  + F_{EDL} =  \frac{1}{{\sigma k_B T}}\int_{ - h}^{{\rm{ - }}h + H} {\left[ { - \frac{{\varepsilon _0 \varepsilon _p }}{2}\left| {\frac{{d\psi }}{{dx}}} \right|^2  + e\psi \left( {n_ +   - n_ -   + n_{H^ +  }  - n_{OH^ -  } } \right)} \right]dx}  \\ 
 +\frac{1}{{\sigma k_B T}}\int_{ - h + H}^0 {\left[ { - \frac{{\varepsilon _0 \varepsilon _r }}{2}\left| {\frac{{d\psi }}{{dx}}} \right|^2  + e\psi \left( {n_ +   - n_ -   + n_{H^ +  }  - n_{OH^ -  } } \right)} \right]dx} - \frac{1}{{\sigma}}\int_{ - h}^{ - h + H} {\left[ {e\psi n_{A^ -  } \phi } \right]dx}  \\ 
   + \frac{1}{{\sigma b^3 }}\int_{ - h}^0 {\left[ {n_ +  b^3 \ln \left( {n_ +  b^3 } \right) + n_ -  b^3 \ln \left( {n_ -  b^3 } \right) + n_{H^ +  } b^3 \ln \left( {n_{H^ +  } b^3 } \right) + n_{OH^ -  } b^3 \ln \left( {n_{OH^ -  } b^3 } \right)} \right]} dx \\ 
  + \frac{1}{{\sigma a^3 }}\int_{ - h}^0 {\left[ {\left( {1 - n_ +  b^3  - n_ -  b^3  - n_{H^ +  } b^3  - n_{OH^ -  } b^3 } \right)\ln \left( {1 - n_ +  b^3  - n_ -  b^3  - n_{H^ +  } b^3  - n_{OH^ -  } b^3 } \right)} \right]} dx \\ 
 + \frac{1}{{\sigma a^3 }}\int_{ - h}^0 {\left[ { - \mu _ +  n_ +   - \mu _ -  n_ -   - \mu _{H^ +  } n_{H^ +  }  - \mu _{OH^ -  } n_{OH^ -  } } \right]} dx
\label{eq:5}
\end{array}
\end{eqnarray}
where $\psi$  is the electrostatic potential,  $n_i$ and $n_{i,\infty}$  are the number density and bulk number density for ion $i$, where  $i =  \pm , H^ + , OH^ -$. Here $n_{i,\infty }  = 10^3 N_A c_{i,\infty}$, where $c_{i,\infty}$  and $N_A$  are the bulk ionic concentration, and Avogardro number, respectively. 

In order to determine the translational entropy of electrolyte ions, 
we use the same lattice statistics as in \cite%
{andelman_prl_1997, iglic_jp_1996}. In the statistics, each electrolyte ion occupies a cell, the length of each side of a cell is $b$. $\mu_-, \mu_+, \mu_{H^+}, \mu_{OH^-}$ are the chemical potentials of cations, anions, hydrogen ions and hydroxyl ions, respectively.

As in \cite%
{das_sm_2019a}, a pH-responsive polyelectrolyte brush is charged by an acidlike dissociation of HA producing $H^ +$ and $A^-$ ions. $K_a$ is the ionization constant for the acidlike dissociation process. The number density of these  $A^-$  ions is expressed as $n_{A^ - }$. 
Consequently,  $F_{ion}$ can be written as
\begin{equation}
F_{ion} = \frac{k_B T}{{\sigma a^3 }}\int_{ - h}^{ - h + H} {\phi \left[ {\left( {1 - \frac{{n_{A^ -  } }}{\gamma }}\right)\ln\left( {1 - \frac{{n_{A^ -  } }}{\gamma }} \right) + \frac{{n_{A^ -  } }}{\gamma }\ln\left( {\frac{{n_{A^ -  } }}{\gamma }} \right) + \frac{{n_{A^ -  } }}{\gamma }\ln\left( {\frac{n_{H^ {+  ,\infty } }}{{K_a' }}} \right)} \right]dx}
\label{eq:6}.
\end{equation}
where $K_a'  = 10^3 N_A K_a ,n_{H^ +  ,\infty }  = 10^3 N_A c_{H^ +  ,\infty }$. $c_{H^ +  ,\infty }$ is the bulk concentration of hydrogen ions, which can be related with bulk $pH$ or $pH_\infty$  as  $c_{H^ +  ,\infty }  = 10^{ - pH_\infty}$, $pK$ as $K_a=10^{-pK}$ and $\gamma$ 
  is the density of polyelectrolyte chargeable sites.
Substituting Eqs. (\ref{eq:2}), (\ref{eq:3}), (\ref{eq:5}), (\ref{eq:6}) into Eq. (\ref{eq:1}), we can finally get the whole expression for the total free energy as
\begin{eqnarray}
\begin{array}{l}
 \frac{F}{{k_B T}} = \frac{3}{{2pa^2 }}\int_{ - h}^{ - h + H} {g\left( {x'} \right)dx'} \int_{ - h}^{x'} {g\left( {x,x'} \right)dx}  + \frac{1}{{\sigma a^3 }}\int_{ - h}^{ - h + H} {f_{EV} \left[ {\phi \left( x \right)} \right]dx} \\ 
+\frac{1}{{\sigma k_B T}}\int_{ - h}^{{\rm{ - }}h + H} {\left[ { - \frac{{\varepsilon _0 \varepsilon _p }}{2}\left| {\frac{{d\psi }}{{dx}}} \right|^2 } \right]dx}  + \frac{1}{{\sigma k_B T}}\int_{ - h + H}^0 {\left[ { - \frac{{\varepsilon _0 \varepsilon _r }}{2}\left| {\frac{{d\psi }}{{dx}}} \right|^2 } \right]dx}  \\ 
  + \frac{1}{{\sigma k_B T}}\int_{ - h}^{ - h + H} {\left[ {\Delta W\left( {n_ +   + n_ -   + n_{H^ +  }  + n_{OH^ -  } } \right)} \right]dx}  \\ 
  + \frac{1}{{\sigma k_B T}}\int_{ - h}^0 {\left[ {e\psi \left( {n_ +   - n_ -   + n_{H^ +  }  - n_{OH^ -  } } \right)} \right]dx}  - \frac{1}{{\sigma k_B T}}\int_{ - h}^{ - h + H} {\left[ {e\psi n_{A^ -  } \phi } \right]dx}  \\ 
  + \frac{1}{{\sigma b^3 }}\int_{ - h}^0 {\left\{ {n_ +  b^3 \ln \left( {n_ +  b^3 } \right) + n_ -  b^3 \ln \left( {n_ -  b^3 } \right) + n_{H^ +  } b^3 \ln \left( {n_{H^ +  } b^3 } \right) + n_{OH^ -  } b^3 \ln \left( {n_{OH^ -  } b^3 } \right)} \right\}} dx \\ 
  + \frac{1}{{\sigma a^3 }}\int_{ - h}^0 {\left\{ {\left( {1 - n_ +  b^3  - n_ -  b^3  - n_{H^ +  } b^3  - n_{OH^ -  } b^3 } \right)\ln \left( {1 - n_ +  b^3  - n_ -  b^3  - n_{H^ +  } b^3  - n_{OH^ -  } b^3 } \right)} \right\}} dx \\ 
  + \frac{1}{{\sigma a^3 }}\int_{ - h}^{ - h + H} {\phi \left[ {\left( {1 - \frac{{n_{A^ -  } }}{\gamma }} \right)\ln \left( {1 - \frac{{n_{A^ -  } }}{\gamma }} \right) + \frac{{n_{A^ -  } }}{\gamma }\ln \left( {\frac{{n_{A^ -  } }}{\gamma }} \right) + \frac{{n_{A^ -  } }}{\gamma }\ln \left( {\frac{{n_{H^ +  ,\infty } }}{{{K_a}' }}} \right)} \right]dx}  \\ 
  + \frac{1}{{\sigma a^3 }}\int_{ - h}^0 {\left[ { - \mu _ +  n_ +   - \mu _ -  n_ -   - \mu _{H^ +  } n_{H^ +  }  - \mu _{OH^ -  } n_{OH^ -  } } \right]} dx \\ 
\end{array}
\label{eq:7}
\end{eqnarray}
Eq. (\ref{eq:7}) should be minimized using the variational formalism in the presence of the following constraints:
\begin{equation}
N = \int_{ - h}^{x'} {\frac{{dx}}{{E\left( {x,x'} \right)}}} ,
\label{eq:8}
\end{equation}
\begin{equation}
N = \frac{1}{{\sigma a^3 }}\int_{ - h}^{ - h + H} {\phi \left( x \right)dx} ,
\label{eq:9}
\end{equation}
where $N$ is the number of Kuhn monomer in every polyelectrolyte brush molecule.
Additionally, we have $\phi \left( x \right)$ related to the functions $g$ and $E$ as
 \begin{equation}
\phi \left( x \right) = \sigma a^3 \int_{ x}^{ - h + H} {\frac{{g\left( {x'} \right)dx'}}{{E\left( {x,x'} \right)}}}.
\label{eq:10}
\end{equation}
This minimization procedure eventually yields the final set of equations dictating the equilibrium of the system. 
We have only considered  $p=1$ corresponding to the case of the fully flexible polyelectrolyte chains.

The Euler-Lagrange equations for $n_{A^-}, \psi, n_{\pm}, n_{H^+}, n_{OH^-}, E\left(x,x'\right), g\left(x\right)$ lead to the following equations:
\begin{equation}
n_{A^ -  }  = \frac{{K_a' \gamma }}{{K_a'  + n_{H^ +  ,\infty } \exp \left( { - \gamma a^3 \frac{{e\psi }}{{k_B T}}} \right)}},
\label{eq:11}
\end{equation}
\begin{equation}
\varepsilon _0 \varepsilon _r \left( {\frac{{d^2 \psi }}{{dx^2 }}} \right) + e\left( {n_ +   - n_ -   + n_{H^ +  }  - n_{OH^ -  }  - n_{A^ -  } \phi } \right) = 0, - h+H \le x \le  0,
\label{eq:12}
\end{equation}
\begin{equation}
\varepsilon _0 \varepsilon _r \left( {\frac{{d^2 \psi }}{{dx^2 }}} \right) + e\left( {n_ +   - n_ -   + n_{H^ +  }  - n_{OH^ -  } } \right) = 0,  - h + H \le x \le 0,
\label{eq:13}
\end{equation}
In the region of $-h\le x\le -h+H$,
\begin{equation}
n_ \pm   = n_{ \pm ,\infty } \exp \left( { \mp \frac{{e\psi }}{{k_B T}}}-\frac{\Delta W}{k_BT} \right)/D_{in},
\label{eq:26}
\end{equation}
\begin{equation}
n_{H^ +  }  = n_{H^ +, \infty } \exp \left( { - \frac{{e\psi }}{{k_B T}}}-\frac{\Delta W}{k_BT} \right)/D_{in},
\label{eq:27}
\end{equation}
\begin{equation}
n_{OH^ -  }  = n_{OH^ -, \infty } \exp \left( { \frac{{e\psi }}{{k_B T}}}-\frac{\Delta W}{k_BT} \right)/D_{in},
\label{eq:28}
\end{equation}
where $D_{in}=1 - \left( {n_{ + ,\infty }  + n_{H^ +  ,\infty }  + n_{ - ,\infty }  + n_{OH^ -  ,\infty } } \right)b^3  \\
+ b^3 \exp\left(-\frac{\Delta W}{k_BT}\right)\left( {\left( {n_{ + ,\infty }  + n_{H^ +  ,\infty } } \right)\exp \left( { - \frac{{e\psi }}{{k_B T}}} \right) + \left( {n_{ - ,\infty }  + n_{OH^ -  ,\infty } } \right)\exp \left( {\frac{{e\psi }}{{k_B T}}} \right)} \right)$.

In the region of $-h+H\le x \le 0$,
\begin{equation}
n_ \pm   = n_{ \pm ,\infty } \exp \left( { \mp \frac{{e\psi }}{{k_B T}}} \right)/D_{out},
\label{eq:14}
\end{equation}
\begin{equation}
n_{H^ +  }  = n_{H^ +  ,\infty } \exp \left( { - \frac{{e\psi }}{{k_B T}}} \right)/D_{out},
\label{eq:15}
\end{equation}
\begin{equation}
n_{OH^ -  }  = n_{OH^ -  ,\infty } \exp \left( {\frac{{e\psi }}{{k_B T}}} \right)/D_{out},
\label{eq:16}
\end{equation}
where $D_{out}=1 - \left( {n_{ + ,\infty }  + n_{H^ +  ,\infty }  + n_{ - ,\infty }  + n_{OH^ -  ,\infty } } \right)b^3  \\
+ b^3 \left( {\left( {n_{ + ,\infty }  + n_{H^ +  ,\infty } } \right)\exp \left( { - \frac{{e\psi }}{{k_B T}}} \right) + \left( {n_{ - ,\infty }  + n_{OH^ -  ,\infty } } \right)\exp \left( {\frac{{e\psi }}{{k_B T}}} \right)} \right)$.

if $\nu\ne 0$ and $\omega \ne 0$,  the equation for monomer distribution profile is
\begin{equation}
\begin{array}{l}
 \phi \left( x \right) = \frac{v}{3\omega}\left\{\left[1 + \kappa ^2 f\left(x\right) \right]^{1/2}  - 1\right\}, 
 \end{array}
\label{eq:17}
\end{equation}
where $f\left(x\right)=\lambda  - \left( x + h \right)^2  + \beta n_{A^-}\psi - \rho \left( 1 - \frac{n_{A^-}}{\gamma}\right)\ln \left(1 - \frac{n_{A^-}}{\gamma}\right)- \rho \frac{n_{A^-}}{\gamma} \ln \left(\frac{n_{A^-}}{\gamma}\frac{{n_{H^ +  ,\infty } }}{{K_a' }}\right)$.

if $\nu\ne 0$ and $\omega=0$, 
\begin{equation}
\phi \left( x \right) = \frac{1}{{2\nu \rho }}f\left(x\right)
\label{eq:17b}
\end{equation}

if $\nu=0$ and $\omega \ne 0$, 
\begin{equation}
\phi \left( x \right) = \frac{1}{{\sqrt {3\omega\rho} }}{{f\left(x\right)}^{1/2}}
\label{eq:17c}
\end{equation}
 
We can be easily aware of that if $ \nu=0.0$ and $\omega=0.0$, there exists a division by zero in Eqs. (\ref{eq:17}), (\ref{eq:17b}) and (\ref{eq:17c}), it is impossible to calculate the monomer distribution profile.

To solve the issue for $\nu=0.0, \omega=0.0$, in the similar way in \cite%
{zhulina_lang_2011}, we can derive a new formula for $\phi\left(x\right)$.(refer Appendix)
\begin{equation}
E\left( {x,x'} \right) = \frac{\pi }{{2N}}\sqrt {\left( {x' + h} \right)^2  - \left( {x + h} \right)^2 },
\label{eq:18}
\end{equation}
\begin{equation}
\left( {q_{tot} } \right)_{H = H_0 }  = 0,
\label{eq:19}
\end{equation}
\begin{equation}
\left( {q_{tot} } \right)_{H = H_0 }  = \frac{e}{\sigma }\int_{ - h}^0 {\left( {n_ +   - n_ -   + n_{H^ +  }  - n_{OH^ -  }  - \phi n_{A^ -  } } \right)dx},
\label{eq:20}
\end{equation}
\begin{equation}
g\left( x \right) = \frac{{\left( {x + h} \right)}}{{\sigma Na^3 }}\left[ {\frac{{\phi \left( { - h + H} \right)}}{{\sqrt {H^2  - \left( {x + h} \right)^2 } }} - \int_x^{ - h + H} {\frac{{d\phi \left( {x'} \right)}}{{dx'}}\frac{{dx'}}{{\sqrt {\left( {x' + h} \right)^2  - \left( {x + h} \right)^2 } }}} } \right].
\label{eq:21}
\end{equation}

Eq. (\ref{eq:11}) denotes the expanded form of the mass action law. Eq. (\ref{eq:12}),(\ref{eq:13}) provides the electrostatic potential distribution in both $-h \le x \le  -h + H$ and $ -h + H \le x$.

Eqs. (\ref{eq:26}-\ref{eq:28}), Eqs. (\ref{eq:14}-\ref{eq:16}) relate the ion number densities to electrostatic potential and the corresponding bulk number density $n_{i,\infty }$ through the Boltzmann distribution.

Eq. (\ref{eq:17}) provides the monomer distribution profile using virial coefficients $\omega$ and $\nu$ , parameters $\rho  = \frac{{8a^2 N^2 }}{{3\pi ^2 }}$, $\lambda  =  - \lambda _1 \rho$ ($\lambda _1$ is the Lagrange multiplier attributed to the constraint condition Eq. (\ref{eq:9})). $\beta  = \frac{{8N^2 ea^5 }}{{3\pi ^2 k_B T}}$. 

While Eq. (\ref{eq:18}) quantifies the local stretching of the polyelectrolyte brush, Eq. (\ref{eq:19}) is used as the condition for determining the equilibrium brush thickness $H_0$.

Finally, Eq. (\ref{eq:21}) expresses the normalized chain end distribution $g(x)$. 
 
Summarizing, the electrostatic potential and the brush thickness for the system consisting of two interaction pH-responsive polyelectrolyte brushes in a good solvent is self-consistently solved by combining Eqs. (\ref{eq:12}-\ref{eq:21}) and the following boundary condition:
\begin{eqnarray}
\begin{array}{l}
 \left( \psi  \right)_{x = \left( { - h + H} \right)^ -  }  = \left( \psi  \right)_{x = \left( { - h + H} \right)^ +  } , \\ 
 \varepsilon_p \left( {\frac{{d\psi }}{{dx}}} \right)_{x = \left( { - h + H} \right)^ -  }  = \varepsilon_r \left( {\frac{{d\psi }}{{dx}}} \right)_{x = \left( { - h + H} \right)^ +  } , \\ 
 \left( {\frac{{d\psi }}{{dx}}} \right)_{x =  - h}  = 0, \\ 
 \left( {\frac{{d\psi }}{{dx}}} \right)_{x = 0}  = 0. \\ 
\end{array}
\label{eq:22}
\end{eqnarray}

In order to combine and solve the equations obtained above, we apply the folloiwng procedure.

Firstly, we give a guess of $H$ and then provide a initial value of $\lambda$. 

Secondly, we can determine $\psi$ by solving Eq. (\ref{eq:12}) and Eq. (\ref{eq:13}). 

Then, $\lambda$ is self-consistently determined with the help of the normalization condition Eq. (\ref{eq:10}).

Next, to obtain  $\phi\left(x\right), \psi\left(x\right), n_{A^-}\left(x\right),n_\pm,n_{H^+},n_{OH^-}$, we use Eq. (\ref{eq:11}), Eq. (\ref{eq:26}-\ref{eq:28}) and Eq. (\ref{eq:14}-\ref{eq:18}).  Finally, the equilibrium brush height $H_0$ will be self-consistently determined by solving Eq. (\ref{eq:19}) and Eq. (\ref{eq:20}).

The solution provides $\phi, \psi, g\left( x \right), H_0, n_{A^ -  }, n_i \left( {i =  \pm ,H^ +  ,OH^ -  }\right)$ and therefore provides the complete equilibrium description of the system.

Multiplying both sides of Eq. (\ref{eq:13}) by $\frac{d\psi}{dx}$   and rearranging the expression, the expression for the electrostatic interaction $P$ is obtained as follows.
\begin{equation}
{\rm{P = }}\Pi {\rm{ + P}}_{{\rm{bulk}}}  =  - \frac{{\varepsilon _0 \varepsilon _r }}{2}\left({\frac{d\psi}{dx}}\right)^2 + 2n_0 k_B T\cosh \left( {\frac{{e \psi }}{{k_B T}}} \right),
\label{eq:23}
\end{equation}
where $n_0=n_{H^+,\infty}+n_{+,\infty}=n_{OH^-, \infty}+n_{-, \infty}$ and $\Pi$ and ${\rm{P}}_{{\rm{bulk}}}$ are the osmotic pressure and the electrostatic interaction for infinite distance between two polyelectrolyte brushes, respectively.
As the separation between the two polyelectrolyte brushes reaches to the positive infinity, the electrostatic interaction ${\rm{P = P}}_{{\rm{bulk}}}$. Substituting ${\rm{P}}_{{\rm{bulk}}}  = 2n_0 k_B T$ into Eq. (\ref{eq:23}), we get the following expression
\begin{equation}
\Pi  =  - \frac{{\varepsilon _0 \varepsilon _r}}{2}\left({\frac{d\psi}{dx}}\right)^2  + 2n_0 k_B T\left[ {\cosh \left( {\frac{{e_0 \psi}}{{k_B T}}} \right) - 1} \right].
\label{eq:24}
\end{equation}
For the sake of ease, if we consider at the centerline, the expression for osmotic pressure can be represented more concisely 
\begin{equation}
\Pi  = 2n_0 k_B T\left[{\cosh \left( {\frac{{e_0 \psi \left( {x = 0} \right)}}{{k_B T}}} \right) - 1} \right].
\label{eq:25}
\end{equation}
It is important to emphasize that we rely on the assumption that two brushes aren't interpenetrate.
Therefore, we consider only the cases when the distance between the two polyelectrolyte brushes is larger than twice the thickness of polyelectrolyte brushes.

\section{Results and Discussion}
 \begin{figure}
\begin{center}
\includegraphics[width=1\textwidth]{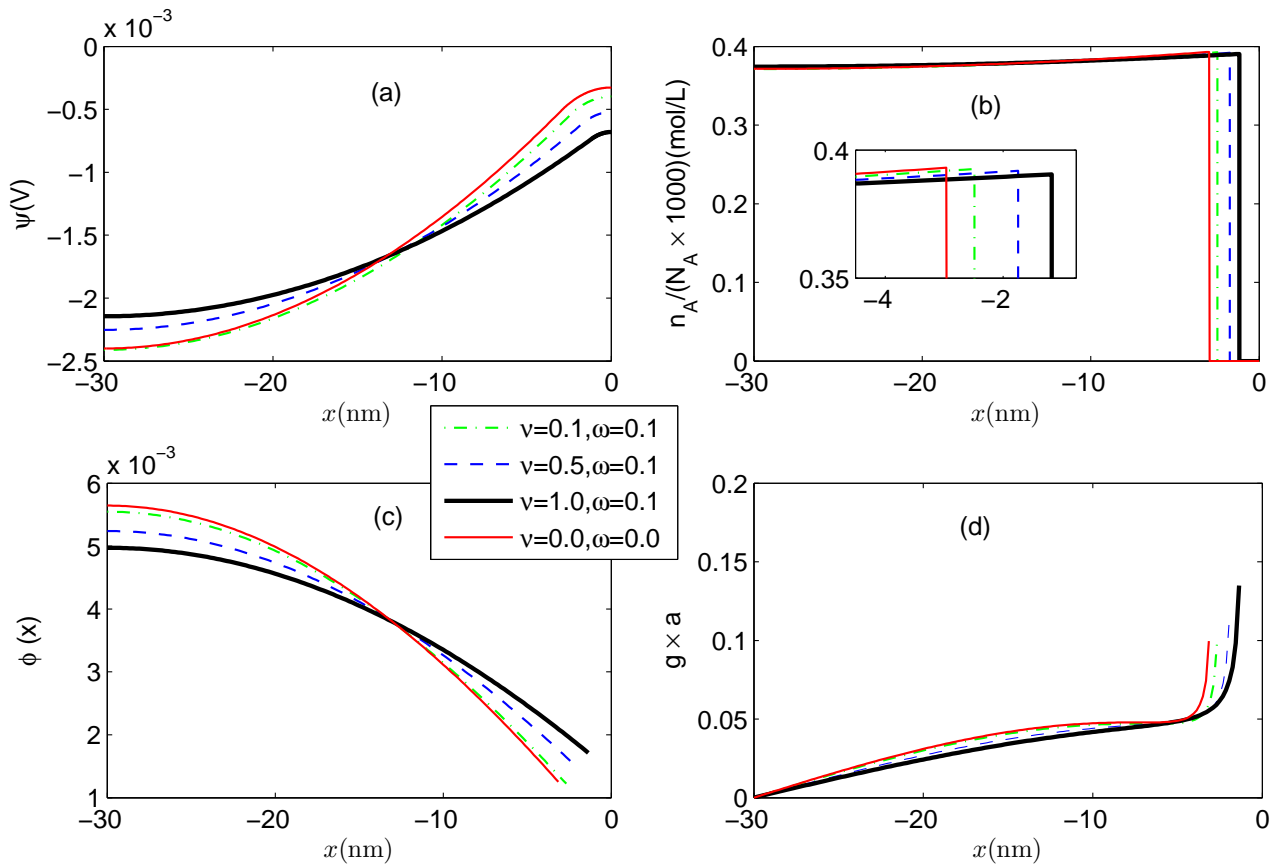}
\caption{(Color online) For two pH-responsive polyelectrolyte brushes, variation of electrostatic potential (a), number density of polyelectrolyte ions (b), monomer distribution profiles (c) , non-dimensional chain end distribution profiles (d) with $x$-coordinate. The case of \cite%
{zhulina_jcp_1997} is the one where $\nu =0$ and $\omega = 0$. $a=1nm, pH=3, pK=3.5, \gamma a^3=1, c_b=0.01mol/L, N=400, l_0=60nm, \zeta=1, b=0, h=30nm$.
}
\label{fig:2}
\end{center}
\end{figure}
 We account for the influence of the excluded volume interaction and the expanded form of the mass action law on pH-responsive brushes to be in a good solvent.
To compare our results with previous studies \cite%
{zhulina_jcp_1997, das_sm_2019a}, we use the same values of parameters $\nu$ and $\omega$ as in the papers.

Fig. \ref{fig:2}(a) shows the electrostatic potential as a function of $x$-coordinate by considering both finite excluded volume interactions and no excluded volume interactions. As in ref \cite%
{das_sm_2019a}, near the surface the magnitude of electrostatic potential increases with excluded volume parameter $\nu$, whereas at the intermediate position ($x=0$) between two polyelectrolyte brushes the electrostatic potential decrease with excluded volume parameter $\nu$. This is explained by the fact that a higher excluded volume interaction leads to a longer thickness of polyelectrolyte brush (see Fig. \ref{eq:2}(b)) and causes a slow change in the electrostatic potential.
Considering the spatial behavior of electrostatic potential, Eqs. (\ref{eq:14}-\ref{eq:16}) provide the fact that the number densities of positive ions and hydrogen ions decrease with the distance from the supporting surface. In contrast, it is clear that the number densities of negative ions and hydroxyl ions increase with the distance from supporting surfaces.
 \begin{figure}
\begin{center}
\includegraphics[width=1\textwidth]{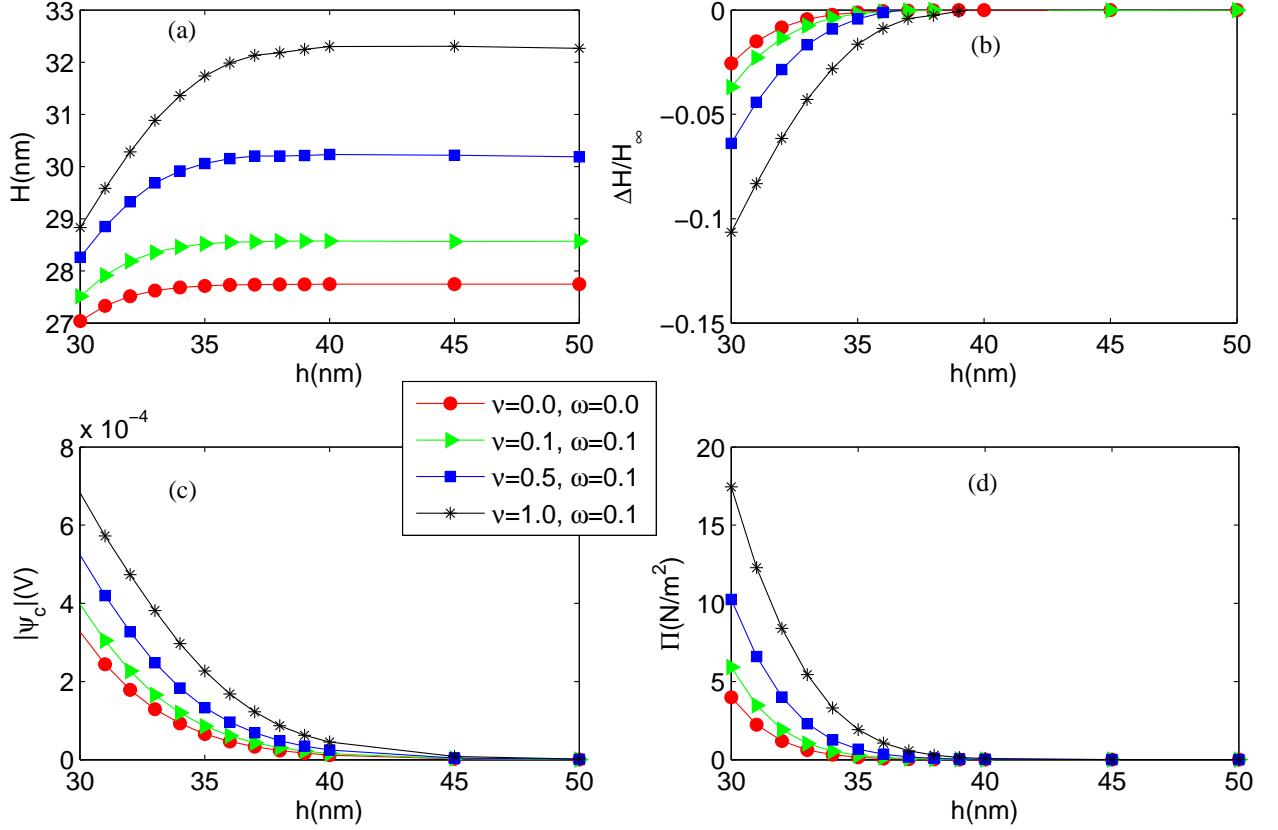}
\caption{(Color online) For two pH-responsive polyelectrolyte brushes, variation of the thickness of polyelectrolyte brush (a), the relative change of thickness of polyelectrolyte brush (b),  the electrostatic potential at the centerlinel (c), the osmotic pressure (d) with the half distance between two pH-responsive polyelectrolyte brushes. The other parameters are the same in Fig. \ref{fig:2}.
}
\label{fig:3}
\end{center}
\end{figure}

Fig. \ref{fig:2}(b) depicts the number density of polyelectrolyte ions as a function of $x$-coordinate. We confirm that a smaller $\nu$ factor provides a shorter polyelectrolyte brush thickness, attributed to excluded volume interaction. Moreover, from the insets we can know that near the end of polyelectrolyte brush, a small $\nu$ provides a larger number density of polyelectrolyte ions. This can be understood by combining Eq. (\ref{eq:11}) and the fact that a small $\nu$ provides a smaller magnitude of electrostatic potential as shown in Fig. \ref{eq:2}(a). 

Fig. \ref{fig:2}(c) displays the dimensionless monomer distribution profile of a polyelectrolyte chain as a function of $x$-coordinate. In the same way in \cite%
{das_sm_2019a}, a small $\nu$ yields a denser monomer concentration near to the wall, and accordingly, in order to ensure a constant N, results in a smaller monomer concentration away from the wall.
 
Fig. 2(d) shows the non-dimensional chain end distribution profiles as a function of the $x$-coordinate. A smaller excluded volume effect provides a higher density of the monomers in the close region from the surface. On the other hand, a large excluded volume effects leads to a flatter distribution of $g$ and ensures a larger $g$ value at farther positions from the surface.
\begin{figure}
\begin{center}
\includegraphics[width=1\textwidth]{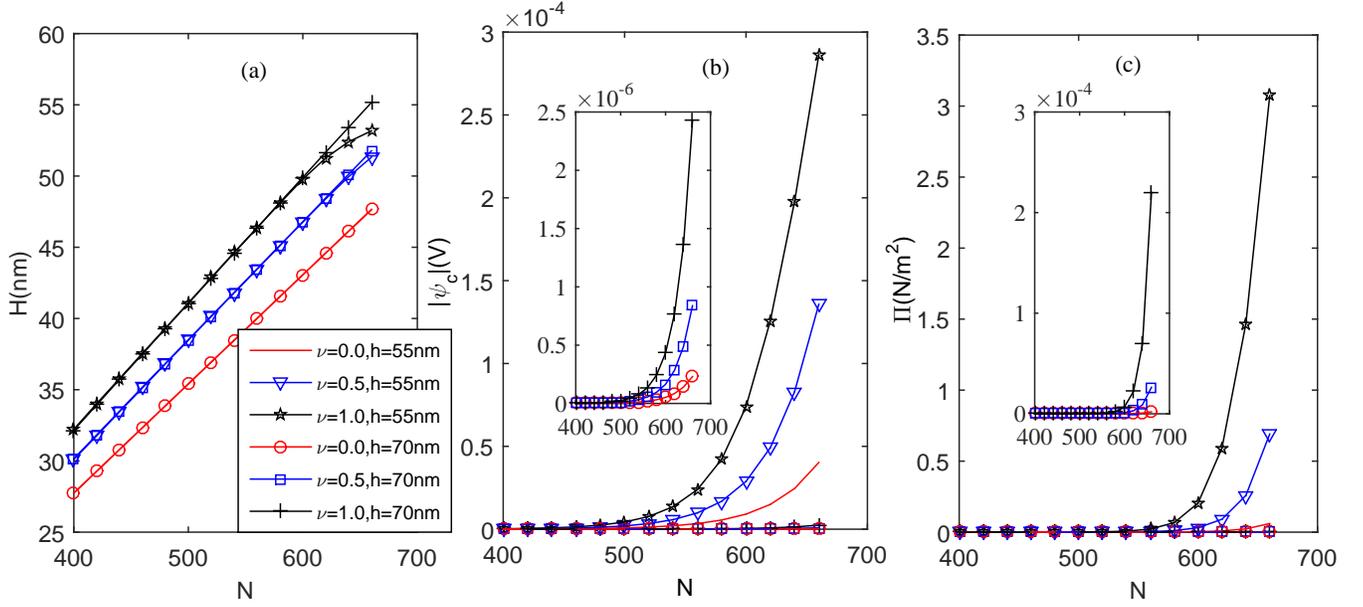}
\caption{(Color online) For two pH-responsive polyelectrolyte brushes, variation of the thickness of polyelectrolyte brush (a), the electrostatic potential at the centerline (b), the osmotic pressure (c) with $N$, the number of Kuhn monomer in every polyelectrolyte brush molecule.
While for $\nu=0.0$, $\omega=0.0$, for the other values of $\nu$, $\omega=0.1$.  The other parameters are the same as in Fig. \ref{fig:2}
}
\label{fig:4}
\end{center}
\end{figure}

Fig. \ref{fig:3}(a) shows the brush thickness as a function of the half separation between two pH-responsive polyelectrolyte brushes for different values of the excluded volume interactions (quantified by different values of $\nu=0.0, \omega=0.0; \nu= 0.1, 0.5, 1, \omega=0.1$). 
We should first indicate that when two pH-responsive polyelectrolyte brushes get close to each other, the equilibrium thickness of polyelectrolyte brush is diminished. i.e. the polyelectrolyte brush contracts. Moreover, such a trend is enhanced with increasing the excluded volume interaction($\nu$). This is attributed to the fact that a larger osmotic pressure due to a higher excluded volume parameter allows the polyelectrolyte brush to contract more strongly.

Fig. \ref{fig:3}(b) shows the relative variation of the brush thickness as a function of the half separation between two pH-responsive polyelectrolyte brushes. The variation is defined as $\Delta H=H-H_{\infty}$,where $H_{\infty}$ is the brush thickness when the distance between two pH-polyelectrolyte brushes is infinite.
The figure displays that at any distance, a large $\nu$ produces a larger relative decrease in brush thickness. The reason is the same as that in Fig.  \ref{fig:3}(a).
In other words, a large excluded volume interaction results in not only a larger absolute contract in brush thickness but also a larger relative contract in brush thickness. This is the first key finding of the present study. 

Fig. \ref{fig:3}(c) depicts the electrostatic potential at the centerline between two pH-responsive polyelectrolyte brushes as a function of the half separation between the brushes for different values of excluded volume parameter ($\nu =0.0, \omega=0.00: \nu=0.1, 0.5, 1, \omega=0.01$). It is shown that a decrease in the distance between two pH-responsive polyelectrolyte brushes results in enhancement in the magnitude of the centerline potential, attributed to the nature of the electrostatic interaction which decreases with the distance between two charged objects. Moreover, a higher excluded volume parameter provides a larger magnitude of electrostatic potential at the centerline between two charged surfaces, while the difference in electrostatic potential for the cases with different excluded volume parameters increases with decreasing the distance between two charged surfaces. This can be explained by following the same reason as in Fig. \ref{fig:2}(a).

Fig. \ref{fig:3}(d) displays the osmotic pressure between two pH-responsive polyelectrolyte brushes as a function of the half separation between the brushes for different values of excluded volume parameter. 
From the formula for osmotic pressure, Eq. (\ref{eq:25}), we can be aware of the fact that the osmotic pressure between two polyelectrolyte brushes decreases with the distance between two surfaces and is enhanced with increasing the excluded volume parameter in the same way as the electrostatic potential at the centerline between the two brushes.

Fig. \ref{fig:4}(a) shows the thickness of polyelectrolyte brush as a function of the number of Kuhn monomer in every polyelectrolyte brush molecule. It displays that the thickness of polyelectrolyte brush increases nearly linearly with the number of Kuhn monomer. Such a behavior can be understood by the fact that the osmotic pressure is very weak and thus the change in the thickness of polyelectrolyte brush due to the osmotic pressure is negligible.

Fig. \ref{fig:4}(b) depicts the electrostatic potential at the centerline between two charged plates as a function of the number of Kuhn monomer in every polyelectrolyte brush molecule. Increasing the number of Kuhn monomer in a polyelectrolyte brush means an increment of the thickness of polyelectrolyte layer and hence yields an enhancement in the magnitude of electrostatic potential at the centerline between two pH-responsive polyelectrolyte brushes.  Due to the same reason as in Fig. \ref{fig:3}(b), an increase in the excluded volume interaction provides an enhancement of the electrostatic interaction.

Fig. \ref{fig:4}(c) displays the osmotic pressure between the two pH-responsive polyelectrolyte brushes as a function of the number of Kuhn monomer in every polyelectrolyte brush molecule. From the formula for the osmotic pressure, the osmotic pressure increases with the number of Kuhn monomer in a polyelectrolyte brush. In addition, the difference in osmotic pressure between the cases with different values of excluded volume parameter is enhanced with the number with Kuhn monomer in a polyelectrolyte brush.
\begin{figure}
\begin{center}
\includegraphics[width=1\textwidth]{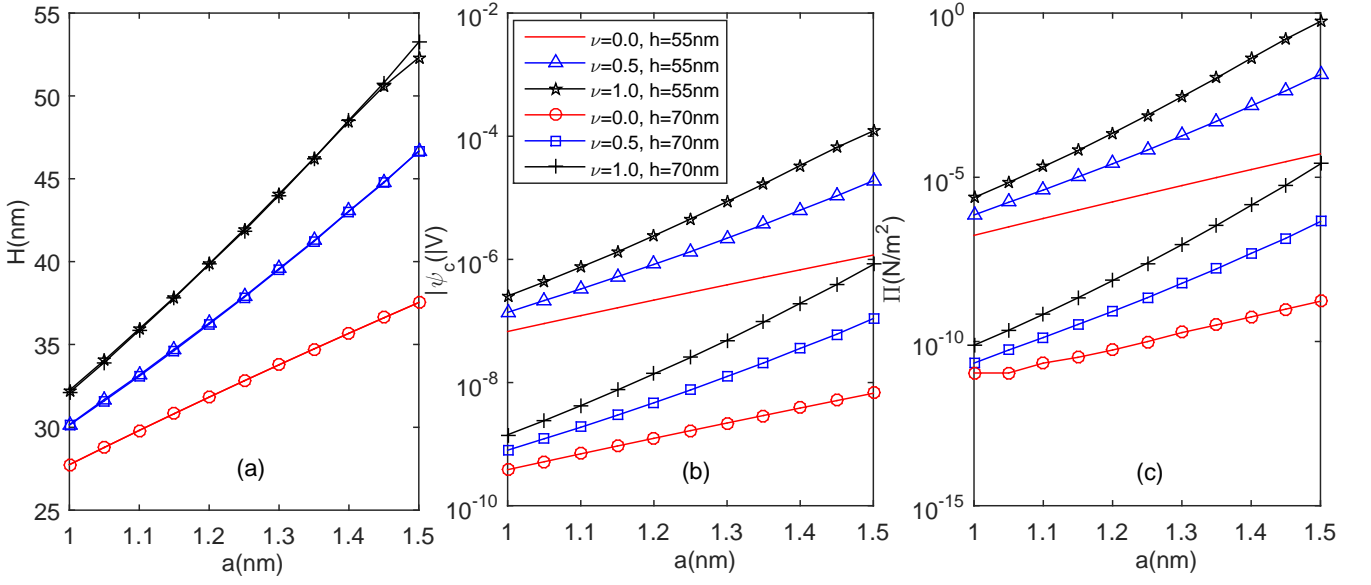}
\caption{(Color online) For two pH-responsive polyelectrolyte brushes, variation of the thickness of polyelectrolyte brush (a), the electrostatic potential at the centerline (b), the osmotic pressure (c) with Kuhn length. While for $\nu=0.0$, $\omega=0.0$, for the other values of $\nu$, $\omega=0.1$. The other parameters are the same as in Fig. \ref{fig:2}.
}
\label{fig:5}
\end{center}
\end{figure} 

Fig.  \ref{fig:5}(a) shows the thickness of polyelectrolyte brush at the centerline between two pH-responsive polyelectrolyte brushes as a function of Kuhn length in every polyelectrolyte brush molecule. An increase of Kuhn length represents an expansion of every polyelectrolyte chain and hence gives an enhancement in the thickness of polyelectrolyte brush. It is noticeable that a larger excluded volume parameter provides a steeper change in the thickness of polyelectrolyte brush. This can be understood by combining the facts that a larger excluded volume parameter provides a stronger excluded volume interaction and that a larger Kuhn length induces a larger deformation of every segment of a polyelectrolyte chain under a given tension. 

Fig.  \ref{fig:5}(b) depicts the electrostatic potential at the centerline between two pH-responsive polyelectrolyte brushes as a function of Kuhn length in every polyelectrolyte brush molecule. The figure shows that the magnitude of electrostatic potential increases with Kuhn length and the difference in the electrostatic potential between different values of excluded volume parameter is enhanced with Kuhn length.
This is attributed to the fact that a large Kuhn length yields a longer thickness of polyelectrolyte brushes (see Fig. \ref{fig:5}(a)) and that a long thickness of polyelectrolyte brush allows the centerline potential to be larger(see Fig. \ref{fig:2}(a)). 
Fig.  \ref{fig:5}(c) displays the osmotic pressure between two pH-responsive polyelectrolyte brushes as a function of Kuhn length in every polyelectrolyte brush molecule. The osmotic pressure increases with Kuhn length and the difference in the osmotic pressure between different values of excluded volume parameter is enhanced with Kuhn length. The fact can be understood by combining the fact that from the definition for osmotic pressure, osmotic pressure increases with the centerline potential and the reason explained in Fig. \ref{fig:5}(b)
 \begin{figure}
\begin{center}
\includegraphics[width=1\textwidth]{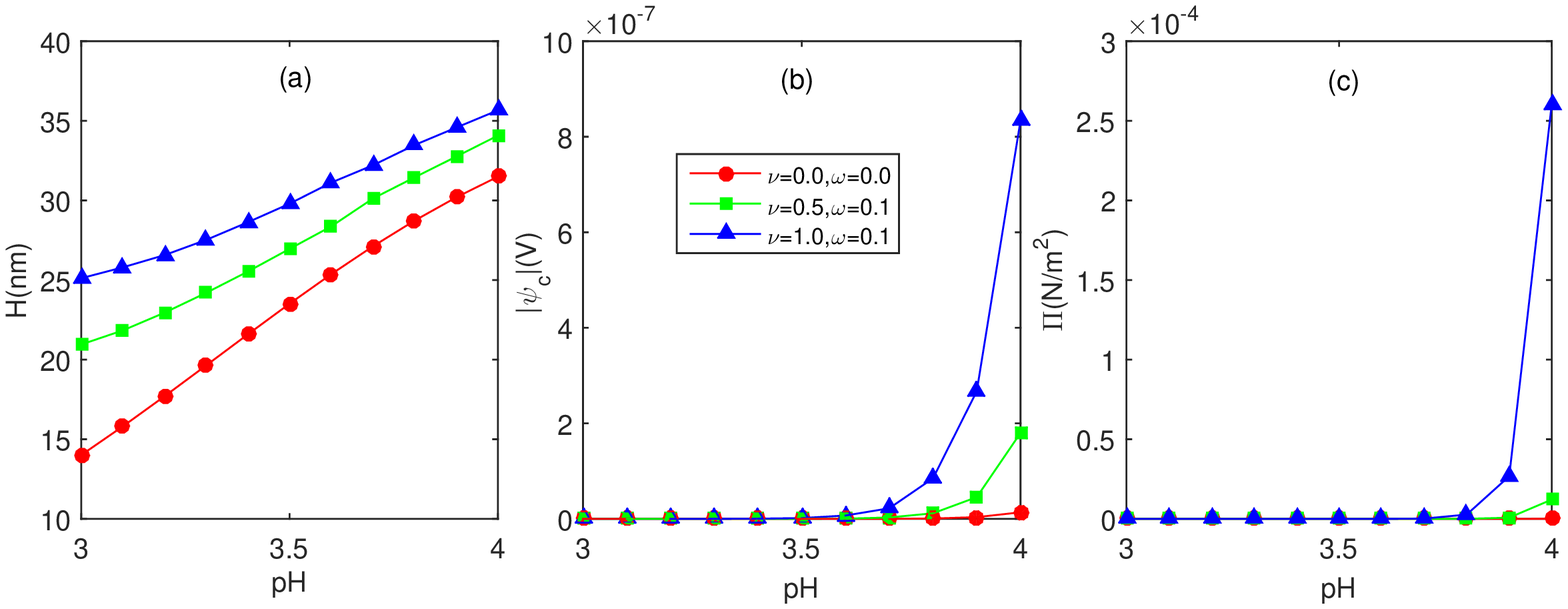}
\caption{(Color online) For two pH-responsive polyelectrolyte brushes, variation of the thickness of polyelectrolyte brushes (a), the electrostatic potential at the centerline (b), the osmotic pressure (c) with pH. $c_b=0.1mol/L, h=40nm$. The other parameters are the same as in Fig. \ref{fig:2}.
}
\label{fig:6}
\end{center}
\end{figure}

Fig. \ref{fig:6}(a) shows the thickness of polyelectrolyte brush as a function of pH. A larger pH (a smaller H+ ion concentration) provides enhanced charging of the polyelectrolyte brushes and hence ensures enhanced counterion-induced osmotic expansion of the brushes causing a larger brush thickness for all values of $\nu$. 
It should be emphasized that as pH increases, the difference in brush thickness between the cases having different values of excluded volume parameter is reduced. This can be explained by the fact that as pH increases, counterion-induced osmotic swelling is enhanced and plays more and more important role in stretching of polyelectrolyte chains compared to excluded volume interaction between monomers in polyelectrolyte chains.

Fig. \ref{fig:6}(b) depicts the electrostatic potential at the centerline between two pH-responsive polyelectrolyte brushes as a function of pH. The figure shows that the magnitude of electrostatic potential increases with pH and the difference in the electrostatic potential between different values of excluded volume parameter is enhanced with pH. Such behavior can be explained by considering the fact that a large pH  allows the thickness of polyelectrolyte brushes to be larger (see Fig. \ref{fig:6}(a)) and a long polyelectrolyte brush yields a lager centerline potential (see Fig. \ref{fig:2}(a)). 
Fig. \ref{fig:6}(c) displays the osmotic pressure between two pH-responsive polyelectrolyte brushes as a function of pH. The osmotic pressure increases with pH and the difference in osmotic pressure between different values of excluded volume parameter is enhanced with pH.  Considering the properties of centerline potential, this behavior can be understood in the similar way as Fig. \ref{fig:5}(c) 
 \begin{figure}
\begin{center}
\includegraphics[width=1\textwidth]{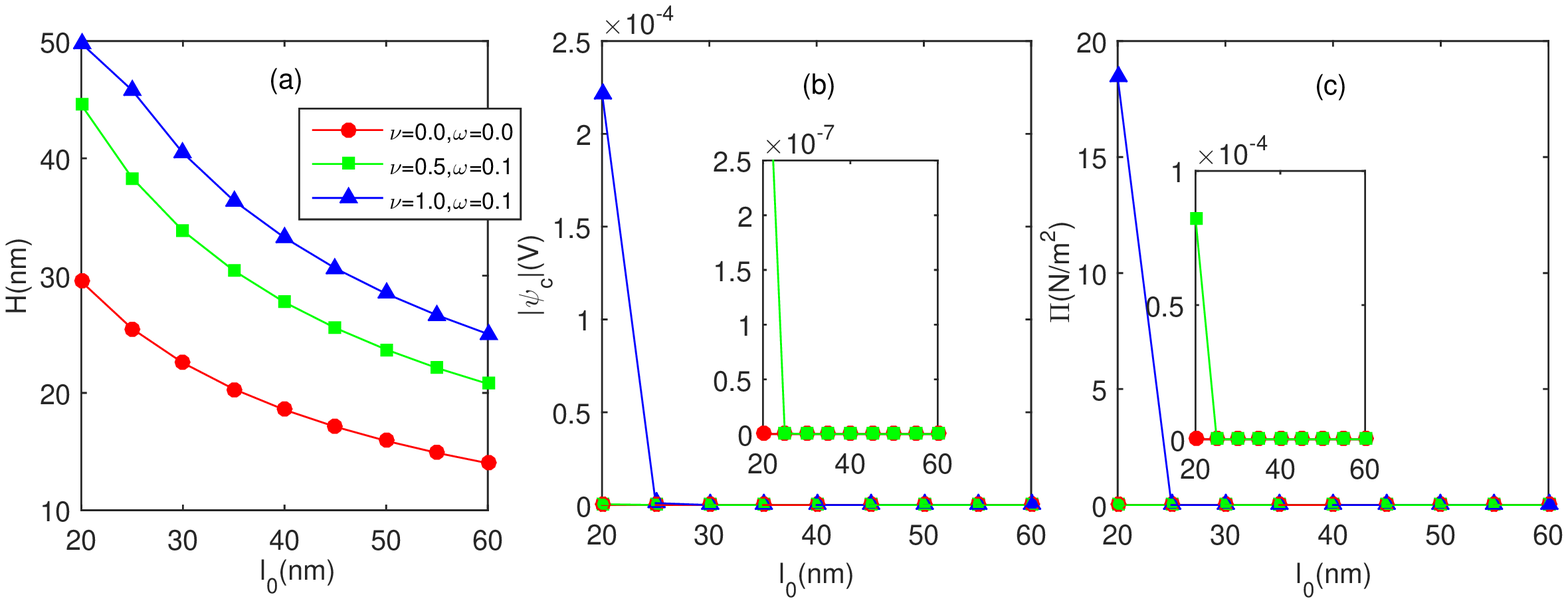}
\caption{(Color online) For two pH-responsive polyelectrolyte brushes, variation of the thickness of polyelectrolyte brush (a), the electrostatic potential at the centerline (b), the osmotic pressure (c) with length $l_0$. $ h=50nm$. The other parameters are the same as in Fig. \ref{fig:6}.}
\label{fig:7}
\end{center}
\end{figure}

Fig. \ref{fig:7}(a) shows the thickness of polyelectrolyte layer as a function of the lateral separation between the adjacent polyelectrolyte brushes. One can first know that the thickness of polyelectrolyte brush decreases with the lateral separation between the adjacent polyelectrolyte brushes.  In fact, from the definition of $l_0$, a larger value of $l_0$ provides a smaller polyelectrolyte brush grafting density and hence a smaller polyelectrolyte ions concentration. Such a smaller polyelectrolyte ion concentration will cause a smaller stretching of polyelectrolyte brushes and provide a smaller thickness of polyelectrolyte brush. 

Fig. \ref{fig:7}(b) depicts the electrostatic potential at the centerline between two pH-responsive polyelectrolyte brushes as a function of the lateral separation between the adjacent polyelectrolyte brushes. The figure shows that the magnitude of electrostatic potential decreases with $l_0$ and the difference in the electrostatic potential between different values of excluded volume parameter is reduced with $l_0$. 
Fig. \ref{fig:7}(c) displays the osmotic pressurebetween two pH-responsive polyelectrolyte brushes as a function of the lateral separation between the adjacent polyelectrolyte brushes. The figure shows that the magnitude of electrostatic potential decreases with $l_0$ and the difference in the osmotic pressure between different values of excluded volume parameter is reduced with $l_0$. The reason for this phenomenon is the same as in Fig. \ref{fig:7}(b).
 \begin{figure}
\begin{center}
\includegraphics[width=1\textwidth]{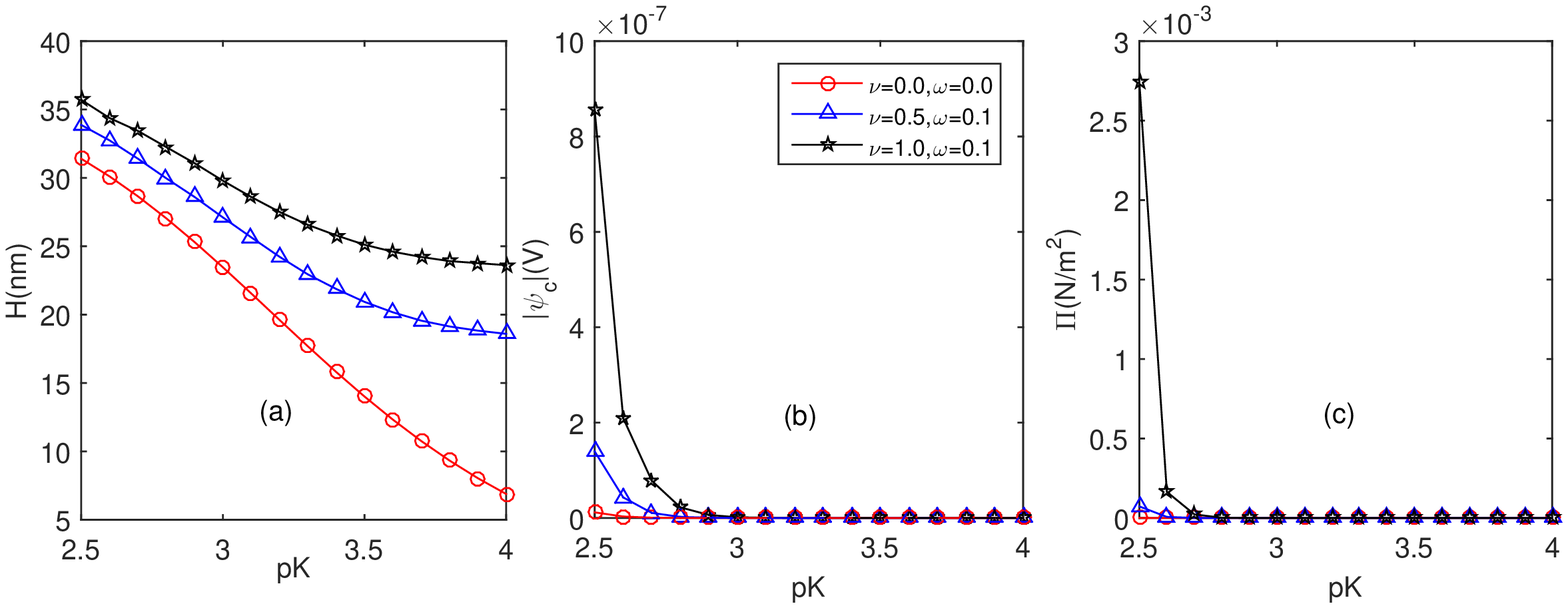}
\caption{(Color online) For two pH-responsive polyelectrolyte brushes, variation of the thickness of polyelectrolyte brush (a), the electrostatic potential at the centerline(b), the osmotic pressure (c) with pK. $h=40nm$. The other parameters are the same as in Fig. \ref{fig:6}.
}
\label{fig:8}
\end{center}
\end{figure}

Fig. \ref{fig:8}(a) shows the thickness of polyelectrolyte brush as a function of pK. A larger pK (a smaller polyelectrolyte ion concentration) provides weakened charging of the polyelectrolyte brushes and hence ensures weakened counterion-induced osmotic swelling of the brushes yielding a smaller brush thickness for all values of $\nu$. 
It should be emphasized that an increase in pK enhances the difference in brush thickness between the cases with different values of excluded volume parameter. This can be understood by the fact that when pK decreases, counterion-induced osmotic swelling increases and plays more and more important role in stretching of polyelectrolyte chains compared to excluded volume interaction between monomers in polyelectrolyte chains.

Fig. \ref{fig:8}(b) depicts the electrostatic potential at the centerline between two pH-responsive polyelectrolyte brushes as a function of pK. The figure shows that the magnitude of electrostatic potential decreases with pK and the difference in the electrostatic potential between different values of excluded volume parameter is reduced with pK.  Fig. \ref{fig:8}(c) displays the osmotic pressure between two pH-responsive polyelectrolyte brushes as a function of pK. The osmotic pressure decreases with pK and the difference in osmotic pressure between different values of excluded volume parameter is reduced with pK. The reason for this phenomenon is the same as in Fig. \ref{fig:8}(b).
 \begin{figure}
\begin{center}
\includegraphics[width=1\textwidth]{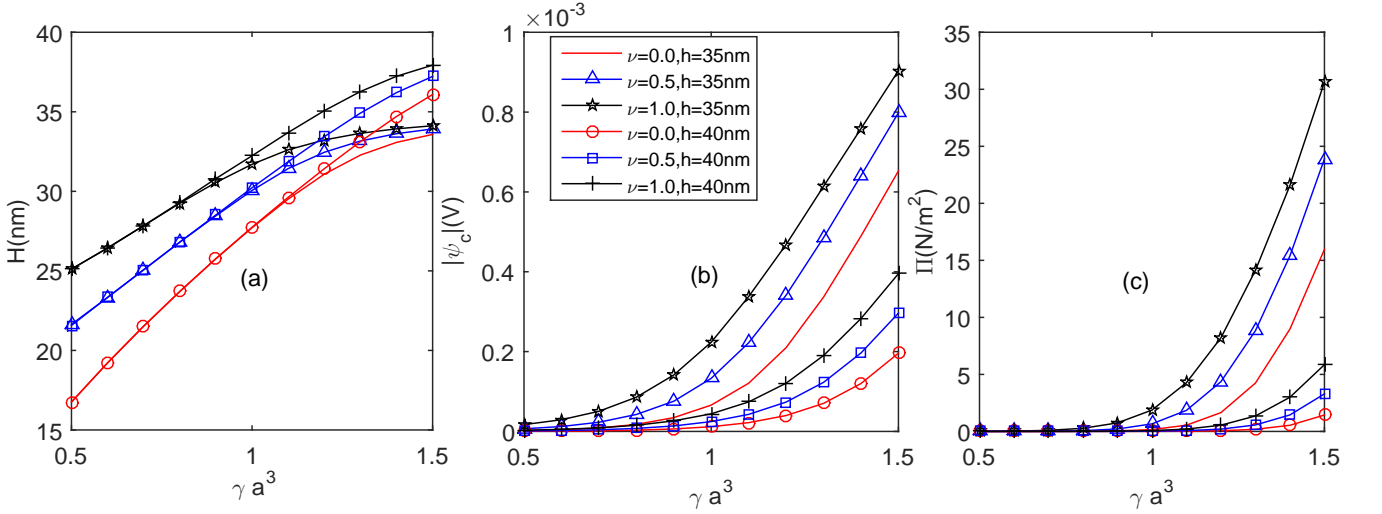}
\caption{(Color online) For two pH-responsive polyelectrolyte brushes, variation of the thickness of polyelectrolyte brush (a), the electrostatic potential at the centerline(b), the osmotic pressure (c) with $\gamma a^3$. While for $\nu=0.0$, $\omega=0.0$, for the other values of $\nu$, $\omega=0.1$. The other parameters are the same as in Fig. \ref{fig:2}.
}
\label{fig:9}
\end{center}
\end{figure} 

Fig. \ref{fig:9}(a) shows thickness of polyelectrolyte brush as a function of $\gamma a^3$. Under a given Kuhn length, a larger $\gamma a^3$ means a stronger density of polyelectrolyte chargeable sites and hence ensures a stronger counterion-induced osmotic swelling of the brushes implying a longer brush thickness for all values of $\nu$.
It should be emphasized that an increase in $\gamma a^3$ diminishes the difference in brush thickness between the cases with different values of excluded volume parameter. This can be proved by understanding that when $\gamma a^3$ increases, counterion-induced osmotic swelling increases and plays more and more important role in stretching of polyelectrolyte chains compared to excluded volume interaction between monomers in polyelectrolyte chains. The reason for this fact is identical to the corresponding one of Fig. \ref{fig:8}(b).

Fig. \ref{fig:9}(b) depicts the electrostatic potential at the centerline between two pH-responsive polyelectrolyte brushes as a function of $\gamma a^3$. The figure shows that the magnitude of electrostatic potential increases with $\gamma a^3$ and the difference in the electrostatic potential between different values of excluded volume parameter enhances with $\gamma a^3$. 
Fig. \ref{fig:9}(c) displays the osmotic pressure between two pH-responsive polyelectrolyte brushes as a function of $\gamma a^3$. The osmotic pressure increases with $\gamma a^3$ and the difference in the electrostatic potential between different values of excluded volume parameter increases with $\gamma a^3$. This phenomenon is attirbuted to the same reason as in Fig. \ref{fig:9}(b).

 \begin{figure}
\begin{center}
\includegraphics[width=1\textwidth]{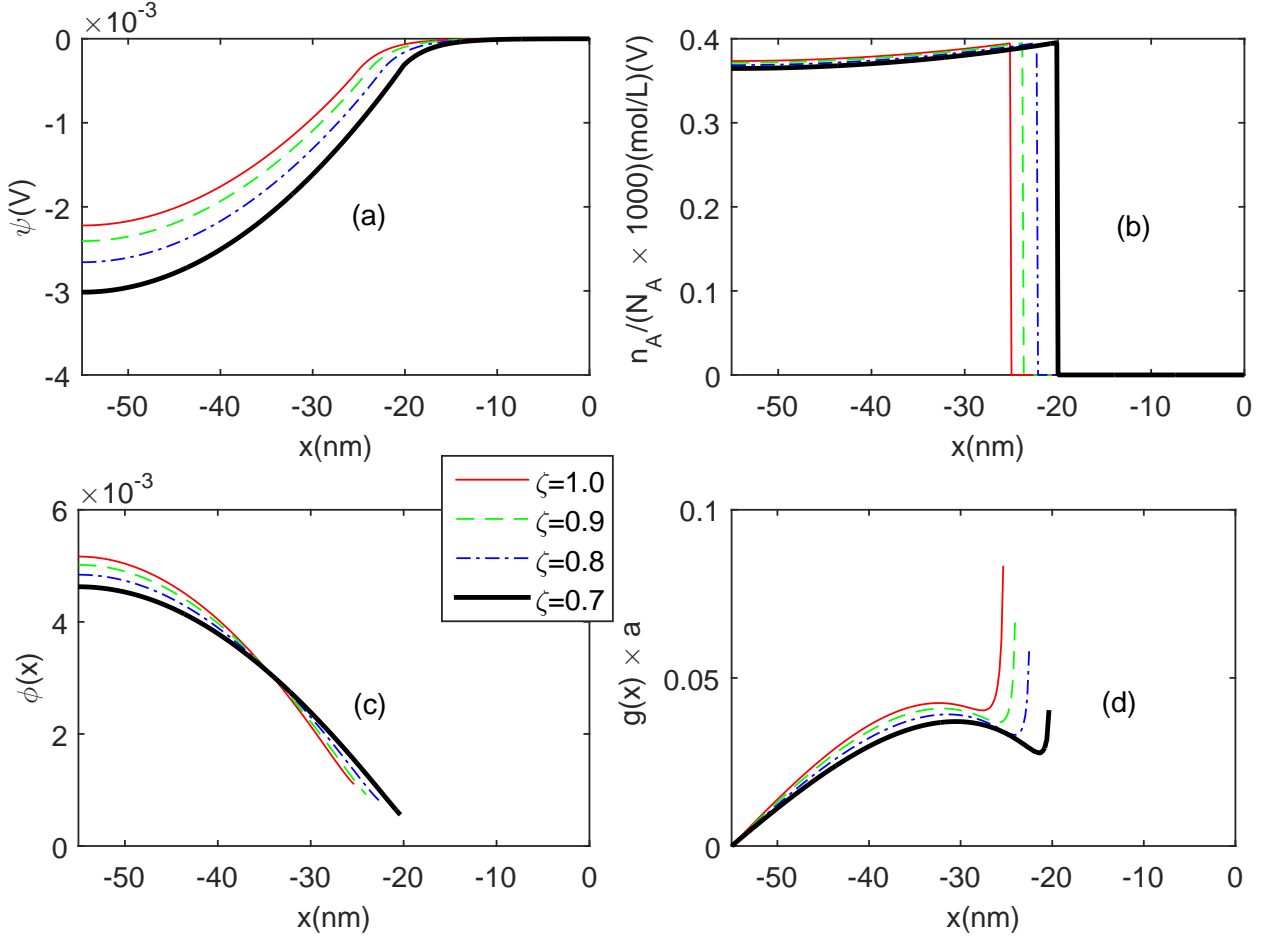}
\caption{(Color online) For two pH-responsive polyelectrolyte brushes, variation of the electrostatic potential (a), concentration of polyelectrolyte ions (b),  dimensionless monomer distribution profile (c), non-dimensional chain end distribution profile (d) with $x$-coordinate. $pH=3, c_b=0.01mol/L, pK=3.5, a=1nm, N=400, l_0=60nm, b=0,\nu=0.5, \omega=0.1, r_0=3.3\times 10^{-10}m$.
}
\label{fig:10}
\end{center}
\end{figure} 

 \begin{figure}
\begin{center}
\includegraphics[width=0.8\textwidth]{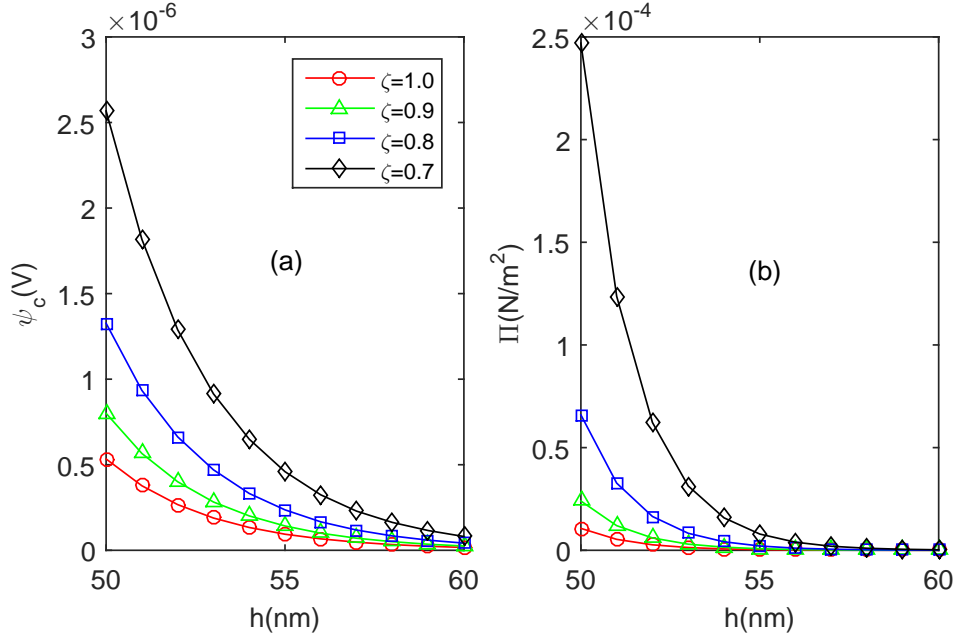}
\caption{(Color online) For two pH-responsive polyelectrolyte brushes, variation of the electrostatic potential at the centerline (b), the osmotic pressure (c) with the half distance between two pH-responsive polyelectrolyte brushes, $h$.  The other parameters are the same as in Fig. \ref{fig:10}.
}
\label{fig:11}
\end{center}
\end{figure} 
 \begin{figure}
\begin{center}
\includegraphics[width=1\textwidth]{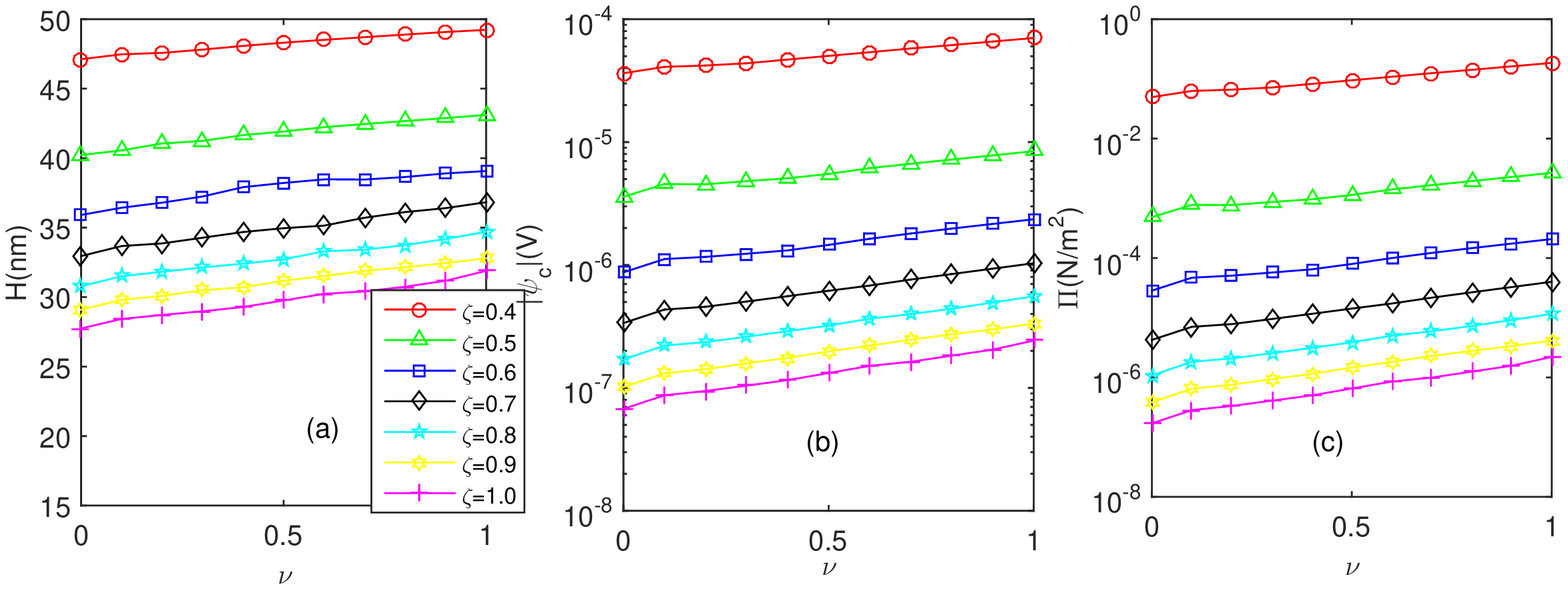}
\caption{(Color online) For two pH-responsive polyelectrolyte brushes, variation of the thickness of polyelectrolyte brush (a), the electrostatic potential at the centerline(b), the osmotic pressure (c) with $\nu$.  $h=55nm$. The other parameters are the same as in Fig. \ref{fig:10}.
}
\label{fig:12}
\end{center}
\end{figure} 

Fig. \ref{fig:10}(a) shows the electrostatic potential as a function of $x$-coordinate for different values of Born energy difference inside polyelectrolyte brush layers.
It turns out from the figure that an increase in Born energy difference (i.e. a decrease in $\zeta$) results in an increase of magnitude of electrostatic potential at all locations. As pointed out in \cite%
{sin_csa_2021, horno_jcis_2003, sadeghi_eleccom_2017, gopmandal_pre_2018, sadeghi_csb_2018, gopmandal_pre_2020, gopmandal_ep_2021, poddar_sm_2016, pandey_csa_2021}, a small values of electric permittivity of polyelectrolyte layer allows Born energy difference to be high. Therefore, overcoming such a region needs more external energy. As a consequence, a smaller permittivity ratio  yields a higher magnitude of electrostatic potential. 

Fig. \ref{fig:10}(b) depicts the concentration of polyelectrolyte ions as a function of $x$-coordinate for different values of Born energy difference inside polyelectrolyte brush layers. Noticeably, we find that when Born energy difference is enhanced, the thickness of polyelectrolyte brush layer is expanded.  This can be explained as follows.
Firstly, an increase in Born energy difference provokes a decrease in number densities of all kinds of electrolyte ions. Thus, the concentration of hydrogen ions becomes lower and accoring to the mass action law, dissociation reaction of polyelectrolyte molecules is enhanced and hence polyelectrolyte ions are more produced. As a results, the stronger the electric repulsion between polyelectrolyte ions, the longer the lengths of polyelectrolyte chains.

Fig. \ref{fig:10}(c) displays the monomer distribtuion profile with $x$-coordinate for different values of Born energy difference inside polyelectrolyte brush layers. From the normalization condition of monomer distribution profiles, near the a core surface the monomer density lowers with increasing Born energy difference(i.e. decrasing $\zeta$), whereas at distant positions from the surface, the monomer density increases with Born energy difference.

Fig. \ref{fig:10}(d) shows non-dimensional chain end distribution profiles with $x$-coordinate for different values of Born energy difference inside polyelectrolyte brush layers. It should be pointed out that inside the brush layer,  a large value of Born energy difference(i.e. a small value of $\zeta$) provides a small value of chain end distribution at any positions inside polyelectrolyte brush layer.

Fig. \ref{fig:11}(a) shows the variation of the electrostatic potential at the centerline between two pH-responsive polyelectrolyte brushes with half separation between the two brushes for different values of Born energy difference inside polyelectrolyte brush layers.
From it, we can see that a decrease in Born energy difference (i.e. an increase in $\zeta$) results in a decrease of magnitude of electrostatic potential at the centerline. This can be understood by following the same reason as in Fig. \ref{fig:10}(a).
Combining  Eq. (\ref{eq:25}) and the above fact, we can understand that as shown in Fig. \ref{fig:11}(b), an increase in Born energy difference provides an increase in osmotic pressure between two pH-responsive polyelectrolyte brushes.

Fig. \ref{fig:12}(a) shows the thickness of pH-responsive polyelectrolyte brush as a function of excluded volume interaction factor $\nu$ for different values of Born energy difference inside polyelectrolyte brush layers.
Combining the the results from Fig. \ref{fig:3}(a) and Fig. \ref{fig:10}(b), both a large excluded volume interaction and a large Born energy difference provide a longer thickness of pH-responsive polyelectrolyte layer. 

We can expect that the electrostatic potential at the centerline and osmotic pressure between two pH-responsive polyelectrolyte brushes increase with both Born energy difference and excluded volume interaction factor(Fig. \ref{fig:12}(b) and Fig. \ref{fig:12}(c)).

Now we should consider ion size effect on the structure and interaction between two pH-responsive polyelectrolyte brushes. Although we have probed ion size effect concretely, it turns out that in weak polyelectrolyte brush, the effect becomes insignificant because the magnitude of the electrostatic potential is not high.

Summarizing, it is elucidated that a large excluded volume interaction yields a long pH-responsive polyelectrolyte brush and a centerline potential and osmotic pressure between two charged surfaces is enhanced for the case with a large excluded volume interaction than a smaller excluded volume interaction.
In particular, as two pH-responsive brushes approach, the absolute and relative decrease in brush thickness increases with excluded volume interaction.
 The result distinguishes our theory from the previous results \cite%
{zhulina_jcp_1997,zhulina_lang_2011,das_sm_2019a}. 

On the other hand, the consideration of density of chargeable sites on a polyelectrolyte molecule involves a change in brush thickness, centerline potential and osmotic pressure. As a result, increasing density of chargeable site on a polyelectrolyte molecule reasonably not only enhances brush thickness, centerline potential and osmotic pressure but also relative decrease of brush thickness when two brushes approach. 

However, the present study does not consider the influence of solvent polarization \cite%
{iglic_bioelec_2012, sin_EA_2015, sin_jcp_2017}, ionic correlation \cite%
{bazant_pre_2012} and surface charge distribution \cite%
{bohinc_jcp_2016} on the interaction between two pH-responsive polyelectrolyte brushes. In fact, because in the paper, the bulk ionic concentration and the magnitude of electrostatic potential is not high, the two effects become trivial. The justification for using the present theory is that we study weak polyelectrolyte brushes, not strong polyelectrolyte brushes. If we consider the case of high salt concentration or high magnitude of electrostatic potential, the ion size, solvent polarization and ionic correlation will be significant. In the future, the effects should be considered by the authors in more detail, in combination with molecular simulations, which will require very expensive cost of computational efforts.

\section{Conclusions}
     In this paper, we predict the behavior of two interacting pH-responsive brush system by using an augmented strong stretching theory considering not only the excluded volume interactions between the polyelectrolyte segments but also a more expanded form of the mass action law valid for $\gamma a^3$.

We unravel that when two interacting pH-responsive polyelectrolyte brushes approach, contraction of the brush thickness is enhanced with the excluded volume interaction and with $\gamma a^3$. 	

We also find that centerline potential and osmotic pressure between two interacting pH-responsive polyelectrolyte brushes decreases with distance, increases with Kuhn length, $\gamma a^3$, pH and number of Kuhn monomer and decreases with pK and the lateral separation between polyelectrolyte chain molecules $l_0$.

 Moreover, we demonstrate that a consideration of Born energy difference enhances thickness of polyelectrolyte brush layer and osmotic pressure between two pH-responsive polyelectrolyte brushes.	

We conclude that simultaneous considerations of excluded volume interaction, density of polyelectrolyte chargeable sites and the Born energy difference are necessary for explaining experiments.

Our study will be useful for explaining physicochemical properties of nanoparticles grafted with pH-responsive brushes to be in a "good" solvent and will be further improved to a more advanced theory dealing with various kinds of polyelectrolytes.

\section{\bf Conflicts of interest}
The authors have no conflicts to disclose.

\section{\bf Data Availavility}
The data that support the findings of this study are available from the corresponding author upon reasonable request.

\appendix
\section{Variation of Free energy for $\nu=0.0$ and $\omega=0.0$}
As already mentioned in Section. II, for case of $\nu=0.0$ and $\omega=0.0$, the formulae for the monomer distribution profile( i.e. Eqs. (\ref{eq:17}, \ref{eq:17b}, \ref{eq:17c})) diverge.

To avoid the difficulty, here, we derive a different solving method for the case without excluded volume interaction($\nu=0.0$ and $\omega=0.0$).
We consider the case accounting for the expanded form of the mass action law and Born energy difference, but not excluded volume interaction between monomers on polyelectrolyte chains. 
 
If we neglect the two effects, the theory leads to ones of \cite%
{zhulina_lang_2011}.

The total free energy functional ($F$) of the PE brush system consists of the elastic ($F_{els}$), electrostatic ($F_{elec}$), and ionization ($F_{ion}$) free energies of polyelectrolyte brush, the electrostatic energy of the EDL ($F_{EDL}$) induced by this brush and the Born energy difference($F_{Born}$). 
\begin{equation}
F = F_{els}  + F_{elec}  + F_{ion}  + F_{EDL}  + F_{Born}. 
\label{eq:A1}
\end{equation}

While $F_{els}$ is given by Eq. (\ref{eq:4}), $F_{elec}  + F_{EDL}$ is written as
\begin{eqnarray}
\begin{array}{l}
 \frac{{F_{elec}  + F_{EDL} }}{{k_B T}} = \frac{1}{{\sigma k_B T}}\int_{ - h}^{{\rm{ - }}h + H} {\left[ { - \frac{{\varepsilon _0 \varepsilon _p }}{2}\left| {\frac{{d\psi }}{{dx}}} \right|^2  + e\psi \left( {n_ +   - n_ -   + n_{H^ +  }  - n_{OH^ -  } } \right) + \Delta W\left( {n_ +   + n_ -   + n_{H^ +  }  + n_{OH^ -  } } \right)} \right]dx}  \\ 
 +\frac{1}{{\sigma k_B T}}\int_{ - h + H}^0 {\left[ { - \frac{{\varepsilon _0 \varepsilon _r }}{2}\left| {\frac{{d\psi }}{{dx}}} \right|^2  + e\psi \left( {n_ +   - n_ -   + n_{H^ +  }  - n_{OH^ -  } } \right)} \right]dx}  - \frac{1}{{\sigma k_B T}}\int_{ - h}^{ - h + H} {\left[ {e\psi n_{A^ -  } \phi } \right]dx}  \\ 
  + \frac{1}{\sigma }\int_{ - h}^0 {\left\{ {n_ +  \left[ {\ln \left( {\frac{{n_ +  }}{{n_{ + ,\infty } }}} \right) - 1} \right] + n_ -  \left[ {\ln \left( {\frac{{n_ -  }}{{n_{ - ,\infty } }}} \right) - 1} \right]} \right\}} dx \\ 
  + \frac{1}{\sigma }\int_{ - h}^0 {\left\{ {n_{H^ +  } \left[ {\ln \left( {\frac{{n_{H^ +  } }}{{n_{H^ +  ,\infty } }}} \right) - 1} \right] + n_{OH^ +  } \left[ {\ln \left( {\frac{{n_{OH^ +  } }}{{n_{OH^ +  ,\infty } }}} \right) - 1} \right]}  + \left( {n_{ + ,\infty }  + n_{ - ,\infty }  + n_{H^ +  ,\infty }  + n_{OH^ -  ,\infty } } \right)
 \right\}dx}  \\ 
 \end{array}
\label{eq:A4}
\end{eqnarray}

$F_{ion}$ and $F_{Born}$ are written as Eq. (\ref{eq:7}) and Eq. (\ref{eq:A5}), respectively.
Substituting Eqs.(\ref{eq:2}), Eq. (\ref{eq:A4}) and Eq. (\ref{eq:A5}) into Eq. (\ref{eq:A1}), the free energy is obtained as follows.
\begin{equation}
\begin{array}{l}
 \frac{F}{{k_B T}} = \frac{3}{{2pa^2 }}\int_{ - h}^{ - h + H} {g\left( {x'} \right)dx'} \int_{ - h}^{x'} {g\left( {x,x'} \right)dx}  +  \\ 
 \frac{1}{{\sigma k_B T}}\int_{ - h}^{{\rm{ - }}h + H} {\left[ { - \frac{{\varepsilon _0 \varepsilon _p }}{2}\left| {\frac{{d\psi }}{{dx}}} \right|^2 } \right]dx}  + \frac{1}{{\sigma k_B T}}\int_{ - h + H}^0 {\left[ { - \frac{{\varepsilon _0 \varepsilon _r }}{2}\left| {\frac{{d\psi }}{{dx}}} \right|^2 } \right]dx}  \\ 
  + \frac{1}{{\sigma k_B T}}\int_{ - h}^{ - h + H} {\left[ {\Delta W\left( {n_ +   + n_ -   + n_{H^ +  }  + n_{OH^ -  } } \right)} \right]dx}  \\ 
  + \frac{1}{{\sigma k_B T}}\int_{ - h}^0 {\left[ {e\psi \left( {n_ +   - n_ -   + n_{H^ +  }  - n_{OH^ -  } } \right)} \right]dx}  - \frac{1}{{\sigma k_B T}}\int_{ - h}^{ - h + H} {\left[ {e\psi n_{A^ -  } \phi } \right]dx}  \\ 
  + \frac{1}{\sigma }\int_{ - h}^0 {\left\{ {n_ +  \left[ {\ln \left( {\frac{{n_ +  }}{{n_{ + ,\infty } }}} \right) - 1} \right] + n_ -  \left[ {\ln \left( {\frac{{n_ -  }}{{n_{ - ,\infty } }}} \right) - 1} \right]} \right\}} dx \\ 
  + \frac{1}{\sigma }\int_{ - h}^0 {\left\{ {n_{H^ +  } \left[ {\ln \left( {\frac{{n_{H^ +  } }}{{n_{H^ +  ,\infty } }}} \right) - 1} \right] + n_{OH^ +  } \left[ {\ln \left( {\frac{{n_{OH^ +  } }}{{n_{OH^ +  ,\infty } }}} \right) - 1} \right]} \right\}dx}  \\ 
  + \frac{1}{{\sigma a^3 }}\int\limits_{ - h}^{ - h + H} {\phi \left( x \right)\left[ {\left( {1 - \frac{{n_{A^ -  } }}{\gamma }} \right)\ln \left( {1 - \frac{{n_{A^ -  } }}{\gamma }} \right) + \frac{{n_{A^ -  } }}{\gamma }\left( {\ln \frac{{n_{A^ -  } }}{\gamma } - \ln \frac{{K_a '}}{{n_{H^ +  ,\infty } }}} \right)} \right]}  \\ 
 \end{array}
\label{eq:A7}
\end{equation}
Eq. (\ref{eq:A7}) should be minimized by performing the variation under the constraints of Eq. (\ref{eq:9}, \ref{eq:10},\ref{eq:11}).
\begin{equation}
\begin{array}{l}
 \delta \left( {\frac{{F_{els} }}{{k_B T}}} \right) = \frac{3}{{2a^2 }}\int\limits_{ - h}^{ - h + x} {\delta g\left( {x'} \right)dx'\int\limits_{ - h}^{x'} {E\left( {x ,x'} \right)dx} }  \\ 
  + \frac{3}{{2a^2 }}\int\limits_{ - h}^{ - h + x} {g\left( {x'} \right)dx'\int\limits_{ - h}^{x'} {\delta E\left( {x,x'} \right)dx} }  + \lambda \int\limits_{ - h}^{ - h + H} {\delta \phi \left( x \right)dx}  \\ 
  - \int\limits_{ - h}^{ - h + H} {\lambda _1 \left( x \right)dx\int\limits_0^x {\frac{{\delta E\left( {x,x'} \right)dx'}}{{E^2 \left( {x,x'} \right)}}} }  \\ 
 \end{array}
\label{eq:A8}
\end{equation}
where
$\delta \phi \left( x \right) = \sigma a^3 \int\limits_x^{ - h + H} {dx'\left( {\frac{{\delta g\left( {x'} \right)}}{{E\left( {x,x'} \right)}} - \frac{{g\left( {x'} \right)\delta E\left( {x,x'} \right)}}{{E^2 \left( {x,x'} \right)}}} \right)}$.
\begin{equation}
\begin{array}{l}
 \delta \left( {\frac{{F_{ion} }}{{k_B T}}} \right) = \frac{1}{{\sigma a^3 }}\int\limits_{ - h}^{ - h + H} {\delta \phi \left( x \right)\left[ {\ln \left( {1 - \frac{{n_{A^ -  } }}{\gamma }} \right) + \frac{{n_{A^ -  } }}{\gamma }\left( {\ln \frac{{\frac{{n_{A^ -  } }}{\gamma }n_{H^ +  ,\infty } }}{{1 - \frac{{n_{A^ -  } }}{\gamma }}} - \ln K_a '} \right)} \right]} dx \\ 
  + \frac{1}{{\sigma a^3 }}\int\limits_{ - h}^{ - h + H} {\frac{{\phi \left( x \right)}}{\gamma }\left[ {\ln \left( {\frac{{\frac{{n_{A^ -  } }}{\gamma }}}{{1 - \frac{{n_{A^ -  } }}{\gamma }}}} \right) - \ln \frac{{K_a '}}{{n_{H^ +  ,\infty } }}} \right]} dx\delta n_{A^ -  }.
 \end{array}
\label{eq:A9}
\end{equation}
The Euler-Lagrange equation for $n_{A^{-}}$  leads to the following equation.
 \begin{equation}
\begin{array}{l}
  - \frac{{e\psi \phi }}{{k_B T}} + \frac{{\phi \left( x \right)}}{{\gamma a^3 }}\left[ {\ln \left( {\frac{{\frac{{n_{A^ -  } }}{\gamma }}}{{1 - \frac{{n_{A^ -  } }}{\gamma }}}} \right) - \ln \frac{{K_a '}}{{n_{H^ +  ,\infty } }}} \right] = 0, \\ 
 \ln \left( {\frac{{\frac{{n_{A^ -  } }}{\gamma }}}{{1 - \frac{{n_{A^ -  } }}{\gamma }}}} \right) - \ln \frac{{K_a '}}{{n_{H^ +  ,\infty } }} = \gamma a^3 \frac{{e\psi }}{{k_B T}}, \\ 
 n_{A^ -  }  = \frac{{K_a '\gamma }}{{K_a ' + n_{H^ +  ,\infty } \exp \left( { - \gamma a^3 \frac{{e\psi }}{{k_B T}}} \right)}}. 
 \end{array}
\label{eq:A15}
\end{equation}
The Euler-Lagrange equation for $\psi$  provides the Poisson equations.
\begin{eqnarray}
\varepsilon _0 \varepsilon _p \left( {\frac{{d^2 \psi }}{{dx^2 }}} \right) + e\left( {n_ +   - n_ -   + n_{H^ +  }  - n_{OH^ -  }  - n_{A^ -  } \phi } \right) = 0,    \left(- h \le x \le  - h + H \right) \\
\varepsilon _0 \varepsilon _r \left( {\frac{{d^2 \psi }}{{dx^2 }}} \right) + e\left( {n_ +   - n_ -   + n_{H^ +  }  - n_{OH^ -  } } \right) = 0,      \left(- h + H \le x \le 0\right)
\label{eq:A16}
\end{eqnarray}
The Euler-Lagrange equations for $n_{\pm}, n_{H^+}, n_{OH^{-}}$  yield the following number densites of cations, anions, hydrogen ions and hydroxyl ions.

In the region of  $- h \le x \le  - h + H$
\begin{eqnarray}
n_ \pm   = n_{ \pm ,\infty } \exp \left( { \mp \frac{{e\psi }}{{k_B T}} - \frac{{\Delta W}}{{k_B T}}} \right),\\
n_{H^ +  }  = n_{H^ +  ,\infty } \exp \left( { - \frac{{e\psi }}{{k_B T}} - \frac{{\Delta W}}{{k_B T}}} \right),\\
n_{OH^ -  }  = n_{OH^ -  ,\infty } \exp \left( {\frac{{e\psi }}{{k_B T}}{\rm{ - }}\frac{{\Delta W}}{{k_B T}}} \right).
\label{eq:A17}
\end{eqnarray}
In the region of  $- h+H < x \le 0$
\begin{eqnarray}
n_ \pm   = n_{ \pm ,\infty } \exp \left( \mp \frac{{e\psi }}{{k_B T}}\right),\\
n_{H^ +  }  = n_{H^ +  ,\infty } \exp \left( - \frac{{e\psi }}{{k_B T}}\right),\\
n_{OH^ -  }  = n_{OH^ -  ,\infty } \exp \left(\frac{{e\psi }}{{k_B T}}\right).
\end{eqnarray}
The Euler-Lagrange equation for $E\left(x, x'\right)$ yields the following expression.
\begin{equation}
\frac{{3g\left( {x'} \right)}}{{2a^2 }} - \left( {\lambda  + \ln \left( {1 - \frac{{n_{A - } }}{\gamma }} \right)} \right)\frac{{g\left( {x'} \right)}}{{E^2 \left( {x,x'} \right)}} - \frac{{\lambda _1 \left( {x'} \right)}}{{E^2 \left( {x,x'} \right)}} = 0.
\label{eq:A20}
\end{equation}
After a brief rearrangement, we get $
E\left( {x, x'} \right) = \sqrt {U_1 \left( {x'} \right) - U_2 \left( {x'} \right)} ,U_1 \left( {x'} \right) = \frac{{2a^2 \lambda _1 \left( {x'} \right)}}{{3g\left( {x'} \right)}},U_2 \left( x \right) =  - \frac{{2a^2 }}{3}\left[ {\lambda  + \ln \left( {1 - \frac{{n_{A^ -  } }}{\gamma }} \right)} \right]$.

Tension should be vanished at the free end of the brush 
 \begin{equation}
E\left( {x',x'} \right) = 0,  x'>0.
\label{eq:A21}
\end{equation}
one finds  $U_1 \left( x \right) \equiv U_2 \left( x \right) \equiv U\left( x \right)$.

By using the normalization condition Eq. (\ref{eq:9}, \ref{eq:10}, \ref{eq:11}), we obtain a well-known result for the Gaussian elasticity of tethered chains.
\begin{equation}
U\left( x \right) = \frac{{\pi ^2 }}{{4N^2 }}\left( {x + h} \right)^2 .
\label{eq:A22}
\end{equation}
Hence, the function of local chain stretching   is given by $E\left( {x,'x} \right) = \frac{\pi }{{2N}}\sqrt {\left( {x' + h} \right)^2  - \left( {x + h} \right)^2 }$.

Combining Eq. (\ref{eq:A20}), Eq. (\ref{eq:A21}) and Eq. (\ref{eq:A22}) and rearranging, we lead to the equation for determining the local number density of the polyelectrolyte ion $A^-$ 
 \begin{equation}
\begin{array}{l}
 \ln \left( {1 - \frac{{n_{A^ -  } }}{\gamma }} \right) =  - \lambda  - \frac{3}{{2a^2 }}U\left( x \right) =  - \lambda  - \frac{{3\pi ^2 }}{{8N^2 a^2 }}\left( {x + h} \right)^2  =  - \lambda  - m^2 \left( {x + h} \right)^2  \\ 
 \frac{{n_{A^ -  } }}{\gamma } =  1 - \exp \left( { - \lambda  - m^2 \left( {x + h} \right)^2 } \right), 
 \end{array}
\label{eq:A23}
\end{equation}
where $\lambda$ is a local Lagrange parameter and  $m^2  = \frac{{3\pi ^2 }}{{8N^2 a^2 }}$.

We introduce the number density of polyelectrolyte ions at the edge of the brush, $n_H  = n_{A^ -  } \left( {x =  - h + H} \right)$, and specify the value of the Lagrange multiplier as 
 \begin{equation}
\begin{array}{l}
 \ln \left( {1 - \frac{{n_H }}{\gamma }} \right) =  - \lambda  - m^2 H^2  \\ 
 \lambda  =  - \ln \left( {1 - \frac{{n_H }}{\gamma }} \right) - m^2 H^2 .
 \end{array}
\label{eq:A24}
\end{equation}
Substituting Eq. (\ref{eq:A24}) into Eq. (\ref{eq:A23}), we obtain the following equation
 \begin{equation}
\frac{{n_{A^ -  } }}{\gamma }=1 - \exp \left( { - \lambda  - m^2 \left( {x + h} \right)^2 } \right) = 1 - \left( {1 - \frac{{n_H }}{\gamma }} \right)\exp \left( {m^2 H^2  - m^2\left( {x + h} \right)^2 } \right).
\label{eq:A25}
\end{equation}
To obtain the electrostatic potential inside the pH-responsive polyelectrolyte brush, we combine Eq. (\ref{eq:A25}) with Eq. (\ref{eq:A15}).
\begin{equation}
\begin{array}{l}
 \frac{{n_{A^ -  } }}{\gamma } = \frac{{K_a '}}{{K_a ' + n_{H^ +  ,\infty } \exp \left( { - \gamma a^3 \frac{{e\psi }}{{k_B T}}} \right)}} = 1 - \left( {1 - \frac{{n_H }}{\gamma }} \right)\exp \left( {m^2 H^2  - m^2\left( {x + h} \right)^2 } \right) \\ 
 \frac{{e\psi }}{{k_B T}} = \frac{1}{{\gamma a^3 }}\left[ {m^2 \left( {\left( {x + h} \right)^2  - H^2 } \right) + \ln \frac{{n_{H^ +  ,\infty } }}{{K_a '\left( {1 - \frac{{n_H }}{\gamma }} \right)}}\left( {1 - \left( {1 - \frac{{n_H }}{\gamma }} \right)\exp \left( {m^2 H^2  - m^2\left( {x + h} \right)^2 } \right)} \right)} \right].
 \end{array}
\label{eq:A30}
\end{equation}

Combining Eq. (\ref{eq:A30}), Eq. (\ref{eq:A16}) and Eq. (\ref{eq:A17}) lead to the following equations for monomer density distribution
\begin{equation}
\begin{array}{l}
 \phi \left( x \right) = \left( {\frac{{\varepsilon _0 \varepsilon _p }}{e}\frac{{d^2 \psi _1 \left( x \right)}}{{dx^2 }} + \left( {n_ +   + n_{H^ +  }  - n_ -   - n_{OH^ -  } } \right)} \right)/n_{A^ -  }  =  \\ 
   = \frac{{2\varepsilon _0 \varepsilon _p }}{{e^2 }}\frac{{m^2 k_B T}}{{\gamma a^3 }}\left[ {\frac{{1 - \left( {1 - \frac{{n_H }}{\gamma }} \right)\exp \left( {m^2 H^2  - m^2 \left( {x + h} \right)^2 } \right)\left( {1 + 2m^2 \left( {x + h} \right)^2 } \right)}}{{\left[ {1 - \left( {1 - \frac{{n_H }}{\gamma }} \right)\exp \left( {m^2 H^2  - m^2 \left( {x + h} \right)^2 } \right)} \right]^2 }}} \right] \\ 
  + \exp \left( { - \frac{{\Delta W}}{{k_B T}}} \right)\left[ {\left( {n_{H^ +  ,\infty }  + n_{ + ,\infty } } \right)\exp \left( { - \frac{{e\psi _1 }}{{k_B T}}} \right) - \left( {n_{OH^ -  ,\infty }  + n_{ - ,\infty } } \right)\exp \left( {\frac{{e\psi _1 }}{{k_B T}}} \right)} \right].
 \end{array}
\label{eq:A31}
\end{equation}
Here, the following equations should be used
\begin{equation}
\exp \left( {\frac{{e\psi _1 }}{{k_B T}}} \right) = \frac{{n_{H^ +  ,\infty } }}{{K_a '\left( {1 - \frac{{n_H }}{\gamma }} \right)}}\left( {1 - \left( {1 - \frac{{n_H }}{\gamma }} \right)\exp \left( {m^2 H^2  - m^2\left( {x + h} \right)^2 } \right)} \right)\exp \left( {\frac{{m^2 \left( {\left( {x + h} \right)^2  - H^2 } \right)}}{{\gamma a^3 }}} \right).
\end{equation}

When we don't consider the Born energy difference and density of polyelectrolyte chargeable sites (i.e.  $\Delta W=0, \gamma a^3=1$), Eq. (\ref{eq:A31}) and Eq. (\ref{eq:A15}) reach to ones of \cite%
{zhulina_lang_2011}.

\nocite{*}
\bibliography{aipsamp}

\end{document}